\documentclass[preprint,aps,floatfix]{revtex4}
\usepackage{graphicx}
\usepackage{amsmath}
\usepackage{epstopdf}  
\begin{document}
\title{
Aging in a Laponite colloidal suspension: 
A Brownian dynamics simulation study 
}
\def\roma{\affiliation{
Dipartimento di Fisica and INFM Udr and SOFT: Complex Dynamics 
in Structured Systems, Universit\`a di Roma ``La Sapienza'', 
Piazzale A. Moro 2, I-00185, Roma, Italy}}
\def\esrf{\affiliation{
European Synchrotron Radiation Facility, B.P. {\em 220}, 
F-{\em 38043} Grenoble, France}}
\author{S. Mossa}\esrf
\author{C. De Michele, F. Sciortino}\roma
\date{\today}
\begin{abstract}
We report Brownian dynamics simulation of the out-of-equilibrium 
dynamics (aging) in a colloidal suspension composed of rigid 
charged disks, one possible model for Laponite, a synthetic clay 
deeply investigated in the last few years by means of various 
experimental techniques. At variance with previous numerical 
investigations, mainly focusing on static structure and equilibrium 
dynamics, we explore the out-of-equilibrium aging dynamics.
We analyze the wave-vector and waiting time dependence of the 
dynamics, focusing on the single-particle and collective density 
fluctuations (intermediate scattering functions), the mean squared 
displacement and the rotational dynamics. Our findings confirm the 
complexity of the out-of-equilibrium dynamical behavior of this class 
of colloidal suspensions and suggest that an arrested disordered state 
driven by a repulsive Yukawa potential, i.e., a Wigner glass can be 
observed in this model.
\end{abstract}
\maketitle
\section{Introduction}
\label{intro:sec}
The out-of-equilibrium dynamical behavior of soft matter---and gels 
in particular---still constitutes an interesting puzzle in condensed 
matter physics~\cite{stadvinphys,cipelletti05,jones02}. 
Indeed, despite the fact that out-of-equilibrium non-ergodic 
soft-materials are ubiquitous in the everyday life, the mechanisms 
governing their dynamics are still unclear. 
An adequate comprehension of the above mechanisms is important
for at least two main general reasons. First, nano-sized particles 
ranging from hundreds of nanometers to microns are currently used 
as building blocks of nanostructered composite materials, therefore 
having strong implications in industrial processing and technology 
development. Second, soft matter systems can help with fundamental 
questions about the nature of condensed matter. Indeed, due the 
fact that interactions can be chemically tuned (strength, range) 
with a high degree of accuracy, they can provide a good 
implementation of particular limits as, for instance, the hard 
sphere purely repulsive potential, and used to test most part of 
the theoretical models developed to address the physics of liquids.

From an experimental point of view, relevant time and length scales 
in colloidal systems are more easily accessible than in the case of 
simple atomic and molecular supercooled liquids and 
glasses~\cite{likos01}. 
Nowadays dynamics can be studied to relatively long time scales and 
average collective dynamics is accessible together with single-particle 
behavior. Different spectroscopies are commonly used, ranging from light 
and X-ray scattering to confocal 
microscopy~\cite{cipelletti05,pontoni03,narayanan06}. 
These techniques allow one to access the instantaneous state of the 
system not only under the form of spectra or correlation functions of the 
observables of interest, but also as direct snapshots (configurations) of 
the system. The direct knowledge of the space position of the particles 
can be directly used to calculate all the relevant quantities or also for 
visual inspection~\cite{courtland02}, sometimes giving a direct physical 
intuition of the mechanisms controlling the dynamics.       

While technological improvement has strongly developed the above experimental 
scattering techniques, theoretical descriptions often lay behind, due to 
difficulties in modeling the interparticle potentials and in a proper description 
of the role of the solvent. Indeed, properties of colloidal systems have often 
been studied by classical Newtonian molecular dynamics simulations, where the 
interaction among particles (either  spherical or anisotropic) are schematized by 
simple isotropic potentials. The above approach completely neglects the interaction 
of the colloidal particles with the solvent where they are dispersed, not to mention 
hydrodynamic effects. 
Important steps have been made in the direction of more realistic descriptions 
implementing Lattice Boltzmann and Fluid Particle Dynamics methods where hydrodynamic 
effects are also considered~\cite{Tan00c,cates04,Horb06}. These approaches are very 
demanding from a computational point of view, strongly limiting both the size of the 
considered systems and the time and length scales accessible by the simulation. 
An intermediate approach is provided by Brownian dynamics simulation (i.e., neglecting 
hydrodynamic interactions) by correctly taking into account the interaction of a 
possibly anisotropic colloidal particle with the solvent and in the presence of a more 
realistic interaction potential among the constituents. This intermediate approach, 
being computationally feasible, could help in understanding out-of-equilibrium 
dynamics on very long time scales. 

In this work we introduce a new Brownian dynamics algorithm and study the 
out-of-equilibrium dynamics of a colloidal suspension formed by strongly 
anisotropic discotic-like molecules, which is believed to be a good model
for {\em Laponite}. Laponite is an interesting material; it is a synthetic 
clay, deeply investigated in the last few years by means of several 
experimental techniques~\cite{Bon98a,Bon02a,Tan04a,Tan05a,Nic00a,Nic01a,Mon05a}. 
This system presents many different open problems, ranging from the correct 
description of the phase 
diagram~\cite{mourchid95,Mon05a,ruzicka04a,ruzicka04b,ruzicka06,bonn99} to 
the out-of-equilibrium and aging 
dynamics~\cite{schosseler,bellour03,cipelletti00}, also under particular 
external conditions~\cite{kaloun05a}. Previous computer simulations have 
focused on the static structure and the role played by the distribution 
of the charges on the platelet~\cite{kutter00,dijkstra97,trizac02}.  
A Brownian dynamics simulation has been presented in Ref.~\cite{odriozola04}, 
with a focus on structure and dynamics at equilibrium. Here we present an 
analysis of the out-of-equilibrium dynamics of a model introduced in 
Ref.~\cite{kutter00}, where a Laponite platelet is modeled by identical 
negatively charged interacting sites, uniformly distributed in a 
two-dimensional disk-like geometry.

The paper is organized as follows. Section~\ref{laponite:sec} contains an 
overview of the experimental phase diagram of Laponite and a description 
of experiments pointing to the existence of an ergodic to non-ergodic phase
transition. Section~\ref{model:sec} is a description of the model we have 
used to describe the interaction between two Laponite platelets, while 
in Section~\ref{algorithm:sec} we briefly recall the principles of 
Brownian dynamics and introduce our algorithm for the integration of the 
equations of motion in the case of strongly anisotropic colloidal molecules. 
Section~\ref{simulation:sec} reports the details of the computer simulations. 
The next four Sections contain a detailed investigation of the long-time 
aging dynamics of the model. Section~\ref{r2:sec} reports a study of the 
mean-squared displacement of the molecules; the one-particle and collective 
dynamics of density fluctuations are included in Sections~\ref{fqt-self:sec} 
and~\ref{fqt-coll:sec}, respectively. Section~\ref{rotations:sec} concludes 
the study of the out-of-equilibrium dynamics with the rotational dynamics 
behavior of the system. Finally, Section~\ref{conclusions:sec} contains a 
discussion of our main results and the conclusions.
\section{Laponite}
\label{laponite:sec}
Laponite is a very interesting material. Not only it is used in several 
industrial processes, but it has also been widely investigated by means 
of the most powerful experimental techniques. It is a synthetic clay, 
formed by very thin cylindrical platelets of radius $r_o\simeq 12.5$~nm 
and thickness $d_o\simeq 1$~nm. The total uniform surface charge is 
$Q\simeq Z e$, with $Z\simeq -700$. Laponite phase diagram has been 
investigated in details in Ref.~\cite{mourchid95}. Recently significant 
modifications to the phase diagram of Ref.~\cite{mourchid95} have been 
proposed by different authors~\cite{Mon05a,ruzicka04a,ruzicka04b,ruzicka06}. 

In a series of papers~\cite{Bon98a,Bon02a,Tan04a,Tan05a}, dynamic arrest in 
Laponite has been interpreted as formation of a {\em Wigner glass}. 
In particular, aging of a glass is related to the presence of cages of 
particles at high concentration, mainly due to repulsive interactions, 
while gelation corresponds to cluster formation due to attraction.
Recently Ruzicka {\em et al.}~\cite{ruzicka04a,ruzicka04b,ruzicka06} 
have performed dynamic light scattering experiments to study aging and 
structural arrest for both low and high concentrations. Measurements 
performed over long periods of time have shown two different routes to 
arrest, depending on the role of attractive short-ranged interactions. 
The situation is reminiscent of the case where competing interactions 
are present, which gives rise to Wigner-type glasses~\cite{Sci04a,mossa04}.  

A different interpretation has been proposed by Nicolai and 
co-workers~\cite{Nic00a,Nic01a,Mon05a} based on a picture in which 
aggregation between the platelets is induced by salt addition, which 
reduces electrostatic repulsion and produces fractal aggregates that will 
eventually form a gel. It was suggested that attractive interaction leads 
to aggregation of the Laponite particles and the formation of a 
{\em house of cards} structure. It was recently shown that the presence 
of positive charges on the rim of the Laponite disks is necessary to 
induce aggregation and gelation~\cite{Mon05a,li05}. These charges were 
neutralized by added pyrophosphate and aggregation and gelation was slowed 
down, even though the resulting ionic strength was increased.  
\section{The atomistic model}
\label{model:sec}
We consider here the realistic interaction model introduced in 
Ref.~\cite{kutter00} and conventionally named {\em ``model A''}, where 
the total negative charge is uniformly distributed over the surface area 
and the possible presence of positive rim charges is neglected. 
In this respect, this work constitutes an effort in the direction of 
quantifying the routes to dynamic arrest driven by repulsive 
interactions which progressively establishes via structural 
reorganization of the system. Future work must address the issue of 
the differences in arrest introduced by the presence of attractions 
mediated by oppositely charged rim particles~\cite{lapomodBdemo}.

Following the authors of Ref.~\cite{kutter00}, each Laponite platelet 
is schematized as a rigid disk, formed by $\nu=61$ charge sites disposed 
on a regular mesh with grid points spaced by $r_o/4$. Each site carries 
a charge $q=Q/\nu\simeq -11.47 e$. The solvent (water in the present case) 
is treated as a continuum of dielectric constant $\epsilon=78$. Three 
among the $\nu$ charge sites are the dynamical sites, i.e., the ones whose 
dynamics is explicitely integrated in time to generate the Brownian 
molecular dynamics trajectory, and have mass $m=747900$~m$_p$ (here $m_p$ 
is the mass of the proton). The remaining interaction sites are mass-less 
and they are only taken into account in the calculation of the forces 
acting on the platelets. Their positions can be trivially calculated at 
any time in terms of the coordinates of the dynamical sites. 

The total interaction energy between two platelets is the sum of $\nu^2$ 
site-site screened Coulomb electrostatic interactions of the Yukawa form. 
The interaction between sites $k$ and $l$ at positions $\vec{r}_k$ and 
$\vec{r}_l$ and pertaining to two different platelets is therefore of 
the DLVO form~\cite{DLVO,kutter00,likos01}:
\begin{equation}
V_{kl}=\frac{q_k q_l e^2}{\epsilon}\frac{e^{-r_{kl}/\lambda_D}}{r_{kl}},
\label{dlvo-eq}
\end{equation}
where $r_{kl}=\| \vec{r}_k - \vec{r}_l \|$ and $\lambda_D$ is the Debye 
screening length of the microions, i. e.,
\begin{equation}
\lambda_D = \sqrt{\frac{\epsilon k_B T}{4\pi (n_+ z_+^2 + n_- z_-^2) e^2}}.
\label{lambda_D-eq}
\end{equation}
In the above equation, $(n_+, z_+)$ and $(n_-,z_-)$ are the concentrations 
and the valences of positive counterions and negative coions coming from 
the platelets and the salt added to the solution. 
\section{Brownian dynamics and the integration algorithm}
\label{algorithm:sec}
In this article we introduce a novel algorithm to integrate the 
Brownian equations of motion which, although neglecting hydrodynamics 
effects, takes into account the presence of the solvent in the dynamics. 
Most Brownian dynamics simulations are based on the original Ermak's 
algorithm~\cite{ermak78} developed for atomic systems, where the 
friction coefficient associated to the damping force and acting 
on spherical objects is isotropic. The Ermak's algorithm has been 
generalized by van Gunsteren and Berendsen~\cite{berend82} to 
the case of molecules formed by spherical centers of interaction, 
for which the choice of an isotropic friction coefficient is 
still appropriate. 
In contrast, in the present case of a strongly anisotropic rigid 
molecule, one should take into account the fact that the platelet 
moves differently in the directions perpendicular and parallel to 
its symmetry axis. 

To the best of our knowledge, the only other attempt to take into 
account in a reasonable fashion the anisotropy of a discotic-like 
particle in a Brownian dynamics computer simulation has been the one 
by Odriozola {\it et al.}~\cite{odriozola04}, based on the integration 
of the Langevin equations of a rigid body, separately for the center
of mass and for the orientation. In the present work we prefer to 
rely on the original work by Ermak, which allows us to find an optimal 
balance between a reasonable realism of the molecular model and an 
acceptable efficiency. This gives us the possibility to follow the 
dynamics of the system on time scales much longer than the ones 
investigated so far.

In our algorithm the viscous damping and the Brownian forces act on each 
of the three sites with mass but the friction coefficient, $\xi$, is 
chosen different in the directions perpendicular ($\xi_\perp$) and 
parallel ($\xi_\parallel$) to the symmetry axis of the platelet. The 
numerical values of the two friction coefficients must be chosen 
properly, in order to reproduce a realistic dynamics for the single 
platelet, as we will describe in details in Section~\ref{simulation:sec}. 
$\xi_\perp$ and $\xi_\parallel$ are used to evaluate the following 
numerical coefficients:
\begin{subequations}
\begin{equation}
c_0^\alpha = e^{-\xi_\alpha \delta t}
\;\;\;\;\;\;\;\;
c_1^\alpha = \frac{1 - c_0^\alpha}{\xi_\alpha \delta t}
\;\;\;\;\;\;\;\; 
c_2^\alpha =\frac{1 - c_1^\alpha}{\xi_\alpha \delta t} 
\end{equation}
\begin{equation}
(\sigma_r^\alpha)^2=\frac{k_B T}{m}\frac{\delta t}{\xi_\alpha}
\left( 2-\frac{3-4 e^{-\xi_\alpha\delta t}+ 
e^{-2 \xi_\alpha \delta t}}{\xi_\alpha \delta t} \right )
\;\;\;\;\;\;\;\;
(\sigma_v^\alpha)^2=\frac{k_B T}{m} (1-e^{-2 \xi_\alpha \delta t})
\label{variance-eq}
\end{equation}
\begin{equation}
 c_{rv}^\alpha = \frac{k_B T}{m}\frac{1}
{\xi_\alpha\sigma_r^\alpha \sigma_v^\alpha} 
(1 - e^{-\xi_\alpha \delta t})^2,
\label{correl-eq}
\end{equation}
\end{subequations}
where $\alpha = \perp, \parallel$ and $\delta t$ is the integration 
time step.

For each dynamical site of mass $m$ of the platelet, and at the time 
$(t+\delta t)$, the integration algorithm can be schematized as follows:
\begin{enumerate}
\item Each pair of components of the vectors
\begin{subequations}
\begin{equation}
\delta{\vec r} =(\delta r_x,\delta r_y,\delta r_z)
\end{equation}
\begin{equation}
\delta{\vec v} =(\delta v_x,\delta v_y,\delta v_z),
\end{equation}
\end{subequations}
is sampled from a bivariate Gaussian distribution with zero mean values, 
variances given by $(\sigma_r^\alpha)^2$ and $(\sigma_v^\alpha)^2$ 
(Eq.~(\ref{variance-eq})) and a correlation coefficient determined by 
$c_{rv}^\alpha$ (Eq.~(\ref{correl-eq})) (see Ref.~\cite{allentildesley}).
With this step we produce a realization of the stochastic process
appropriate to generate Brownian motion
\item Transform position, ${\vec r}(t)$, and velocity, ${\vec v}(t)$  
evaluated in the laboratory-fixed frame at time $t$, to ${\vec r'}(t) = 
(r'_x,r'_y,r'_z)$ and ${\vec v'}(t) = (v_{x}', v_{y}', v_{z}')$ in the 
body (platelet)-fixed reference frame, where the $z$-axis is chosen 
parallel to the platelet symmetry axis 
\begin{subequations}
\begin{equation}
{\vec r'}(t) = {\bf R}\, {\vec r}(t) 
\end{equation}
\begin{equation}
{\vec v'}(t) = {\bf R}\, {\vec v}(t); 
\end{equation}
\end{subequations}
here ${\bf R}$ is the appropriate orthogonal matrix corresponding to the 
change of the reference frame
\item Update positions in the body-fixed reference frame to $(t+\delta t)$ 
and velocities to $(t+\delta t/2)$, according to Ermak's prescription:
\begin{subequations}
\begin{equation}
{r'_\beta}(t+\delta t)={r_\beta}'(t)+ c_0^\alpha v_{\beta}' 
\delta t + c_2^\alpha \frac{F_{\beta}(t)}{m}\delta t^2 + \delta r_{\beta} 
\end{equation}
\begin{equation}
{v}'_{\beta}(t+\frac{\delta t}{2}) = c_0^{\alpha} {v}'_{\beta}(t) + 
(c_1^{\alpha} - c_2^{\alpha}) \frac{F_{\beta}(t)}{m} \delta t+ 
\delta v_{\beta},
\end{equation}
\end{subequations}
where $\alpha=\parallel$ if $\beta=x,y$, and $\alpha=\perp$ if $\beta=z$.
Note that $F_{\beta}(t)$ are the forces evaluated at time $t$ and also 
include the intra-platelet contributions coming from the mass-less
sites
\item Transform positions and velocities to the laboratory-fixed reference 
frame:
\begin{subequations}
\begin{equation}
{\vec r}(t + \delta t) = {\bf R}^{-1} {\vec r'}(t+\delta t)  
\end{equation}
\begin{equation}
{\vec v}(t + \delta t) = {\bf R}^{-1} {\vec v'}(t+\delta t)  
\end{equation}
\end{subequations}
\item Adjust positions and velocities using the RATTLE 
algorithm~\cite{allentildesley}, in order to fulfill the rigid constraints 
(fixed distances) among the three sites with mass.
\item Evaluate the forces $F_{\beta}(t+\delta t)$ acting on the site at 
time $(t+\delta t)$.
\item Update velocities to $t+\delta t$ following Ermak's scheme:
\begin{equation}
v_{\beta}(t+\delta t) = v_{\beta}(t+\frac{\delta t}{2}) + 
c_2^{\alpha} \delta t \frac{F_{\beta}(t+\delta t)}{m},  
\end{equation}
where, again, $\alpha=\parallel$, if $\beta=x,y$ and $\alpha=\perp$, 
if $\beta=z$.
\item Finally, adjust velocities using the RATTLE algorithm to assure zero 
relative velocity among the dynamical sites.
\end{enumerate}
\section{Simulation details}
\label{simulation:sec}
We have simulated a system composed of $N=108$ Laponite platelets of 
radius $r_o=12.5$~nm. The time unit is $t_0=73.6$~ps and the time step 
used to integrate the equation of motions is  $\delta t=0.1\: t_0$. 
The system is initially thermalized at a high temperature, $T=500$~K, 
and at time $t_w=0$, it is instantaneously quenched at the room
temperature, $T=300$~K. $t_w$ is the {\em waiting time}, i.e., the 
elapsed time after the quench, and it is an additional relevant time 
scale for the dynamics of the system, as we will see below. 
The total volume of the system is $V=1.515 \times 10^6$~nm$^3$, which 
corresponds to a volume fraction $\phi=3.5\%$ in the case we assume that 
each platelet occupies a volume equal to $4\pi r_o^2 d_o=1963.5$~nm$^3$. 
The side length of the simulation box is $L=114.84$~nm and cubic 
periodic boundary conditions are used. The Debye screening length 
value chosen for the interaction potential Eq.~(\ref{dlvo-eq}) is 
$\lambda_D=3$~nm, a value which is believed to be appropriate to 
realistic experimental conditions~\cite{kutter00}.
The interaction potential is truncated when the distance between the 
two interacting sites exceeds a cut-off radius $r_c=10$~nm. 

The numerical values of the friction coefficients, $\xi_\perp$ and 
$\xi_\parallel$, defined above have been chosen according to the 
following argument. A Laponite platelet can be assimilated to an 
oblate ellipsoid of semi-axes $(a,b,c)$, with $a=b$ and $a > c$, 
which diffuses in a solvent (water) characterized by a viscosity 
$\eta$. In the case of an oblate ellipsoid, the translational 
diffusion coefficients, $D_\perp^T$ and $D_\parallel^T$---corresponding 
respectively to displacements perpendicular and parallel to the 
platelet surface---, and the corresponding rotational diffusion 
coefficients, $D_\perp^R$ and $D_\parallel^R$---that correspond to 
rotations about axes perpendicular and parallel to the platelet surface 
respectively---, can be evaluated exactly 
(see, for instance, Ref.~\cite{shimizu}). More specifically, assuming 
that: {\em i)} we consider as a solvent pure water, i.e., $\eta=1.002$~cP; 
{\em ii)} the freely diffusing oblate ellipsoid has an aspect ratio 
similar to that of the Laponite platelets, i.e., $a=12.5$~nm and $b=0.5$~nm; 
{\em iii)} the temperature is $T=300$~K; we have chosen the friction 
coefficients $\xi_\perp=0.016$~ps$^{-1}$ and 
$\xi_\parallel=0.063$~ps$^{-1}$. 

As we will see below, following the temperature jump, the simulated system 
does not reach equilibrium on the time scale of the simulation. This implies 
that two-time correlation functions will depend separately on the time 
difference $t$ and the time $t_w$ elapsed from the quench. Hence, average 
over starting times must be substituted with an average over an ensemble 
of several statistically independent initial configurations. We have 
considered $100$ independent configurations of the systems which have been  
equilibrated at high temperature $T=500$~K and, at time $t_w=0$, quenched 
instantaneously to ambient temperature $T=300$ K. We have calculated the 
Brownian dynamics trajectories for each sample for $10^6$ integration time 
steps, corresponding to a total time interval of $7 \mu s$. The total CPU 
time amounts to about $14000$ hours on a farm of $10$ $64$MHz Athlon CPUs. 
The evolution of the $100$ independent systems has been recorded and system 
configurations have been stored logarithmically spaced in time for the 
analysis presented in the following Sections. 

\begin{figure}[h]
\centering
\vspace{0.5cm}

\includegraphics[width=0.50\textwidth]{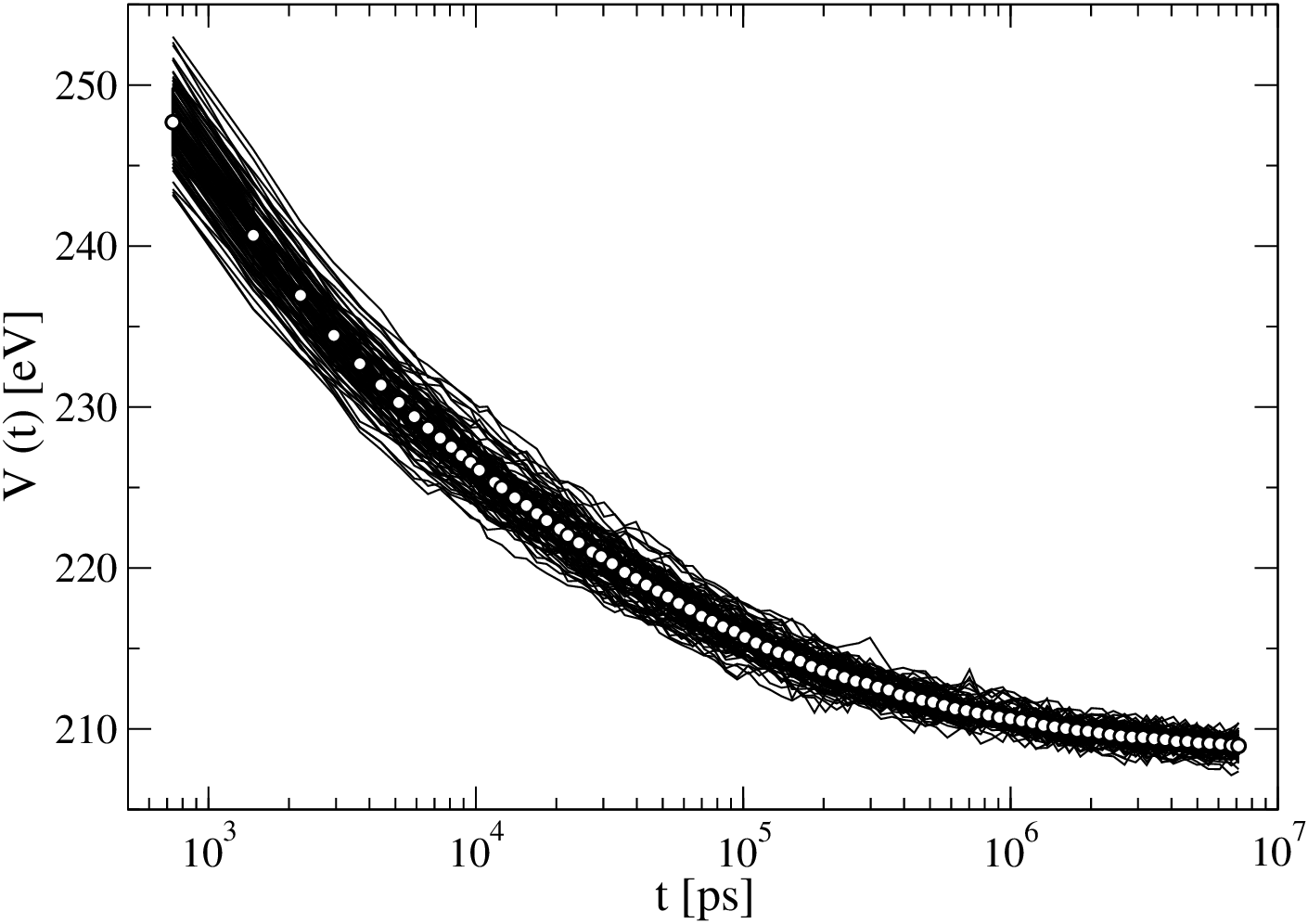}
\caption{}
\label{fig:ene}
\end{figure}

Following the temperature jump from high temperature, the energy of the 
system relaxes but never reaches thermodynamic equilibrium at the lower 
temperature. In Fig.~\ref{fig:ene} we show the energy relaxation for each 
sample (solid lines) and the average value (open circles). Note that, due 
to the repulsive nature of the interaction potential, the total potential 
energy is always positive. The relaxation follows a power law of exponent 
$\gamma\simeq 0.35$ and the equilibrium value (estimated at $E_{\infty} 
\simeq 206.6$~eV) is not reached on the simulation time scale. Hence, 
the system is always out-of-equilibrium, and explores a series of metastable 
states of lower and lower energy. In what follows we will characterize the 
aging dynamics, and show that the slowing down of the dynamics strongly 
depends on the waiting time, $t_w$.

\begin{figure}[h]
\centering
\vspace{0.5cm}

\includegraphics[width=0.50\textwidth]{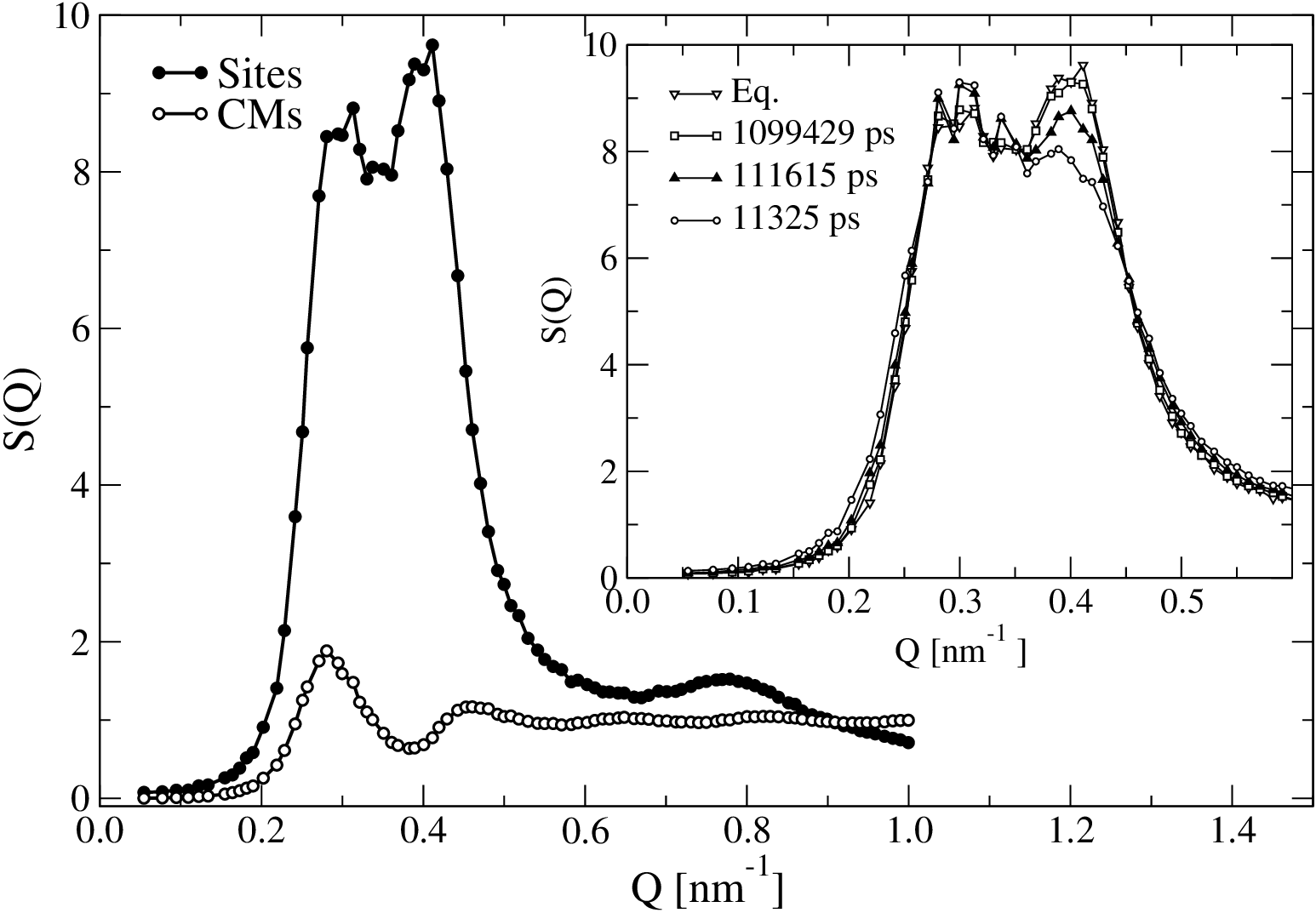}
\caption{}
\label{fig:sk}
\end{figure}

In the main panel of Fig.~\ref{fig:sk} we show the static structure factor, 
$S(Q)=\langle\rho(Q)\rho^*(Q)\rangle$, where 
$\rho(Q)=1/M\sum_i^M \exp(i\, \vec{Q}\cdot \vec{X}_i)$ 
is the density fluctuation of wave-vector $\vec Q$. Here and in the following, 
the coordinates $\vec X_i$ indicate either the center of mass of platelet 
$\vec R_i$ (in this case $M=N$) or the position $\vec r_i$ of all sites 
(in this case $M=61\times N$). The figure shows our results for both centers 
of mass (open symbols) and sites (closed symbols). In particular, latter data 
contain detailed informations on the form factor of the clay platelets and, 
therefore, they are the results most appropriate for a comparison with 
experimental measurements. We note, however, that the smallest wave-vector 
value accessible in our simulation has modulus $Q_m=2\pi/L\simeq 0.055$~nm$^{-1}$, 
obviously far from the values typical of the most popular scattering techniques. 
Each point shown is a spherical average in momentum space with a resolution 
$\Delta Q=0.05$~nm$^{-1}$ and up to $200$ vectors have been considered for 
each value of $Q$. The data have been calculated from the configurations 
produced at times larger than $10^6$~ps and no changes have been found in 
the static structure at later times. Only at much shorter times small 
changes are evident in the structure of the second sub-peak for the case of
interaction sites, as shown in the inset of Fig.~\ref{fig:sk}.

In summary, from the above data it is evident that, following the temperature
jump, the system starts to explore out-of-equilibrium states of lower and lower
energy. During the above process the static structure does not significantly 
change at long times, as also found in the case of molecular glass formers.
This is at strong variance with the long-time structural dynamics---as shown 
by the mean-squared displacement and density fluctuations correlation 
functions---, which show spectacular changes, as we will report in the 
following Sections.  
\section{The mean-squared displacement}
\label{r2:sec}
\begin{figure}[h]
\centering
\vspace{0.5cm}

\includegraphics[width=0.50\textwidth]{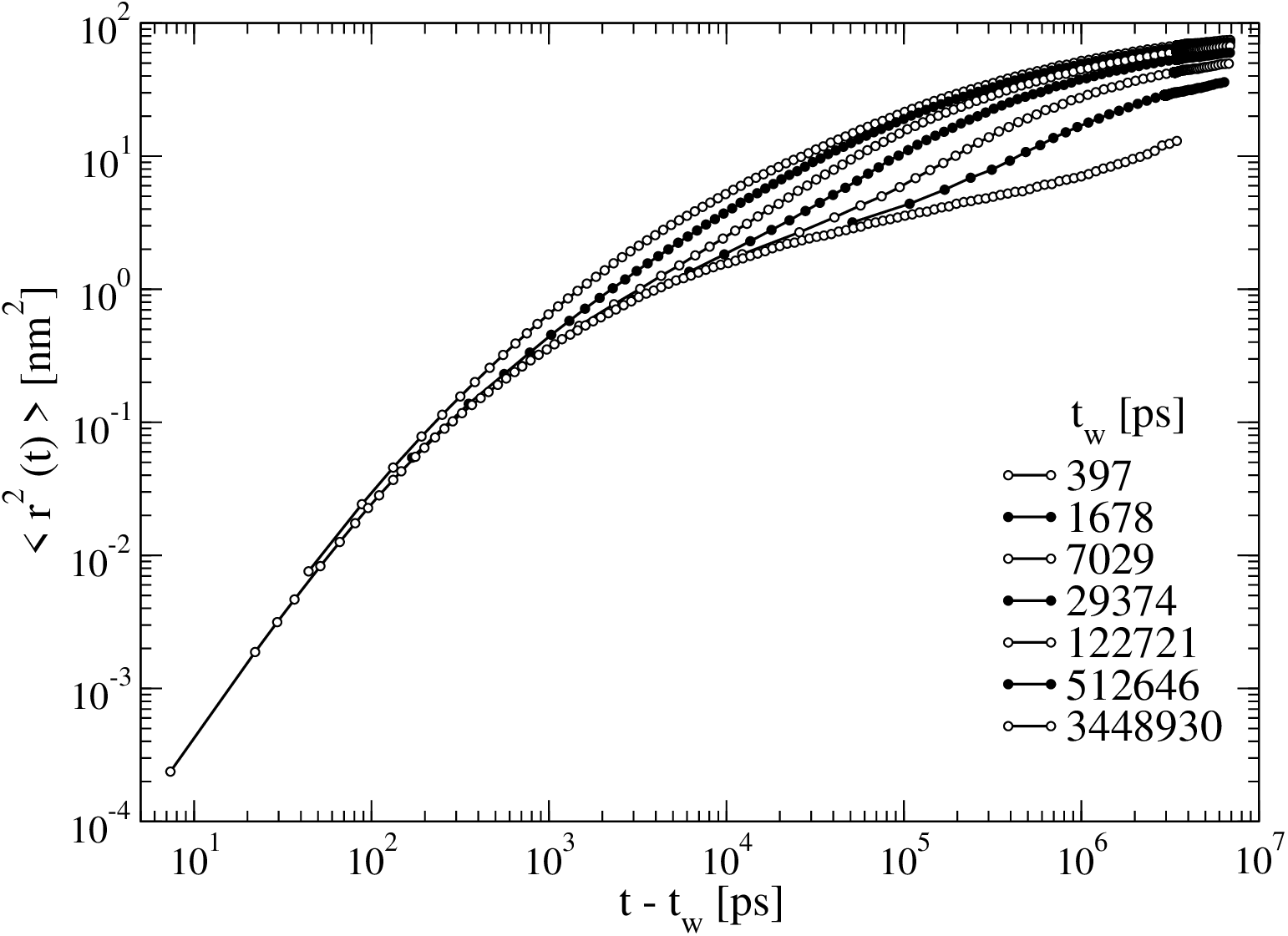}
\caption{}
\label{fig:msd}
\end{figure}

In this Section we analyze the centers of mass mean-squared displacement of 
the center of mass of the platelets,
\begin{equation}
\langle r^2(t,t_w)\rangle=\frac{1}{N}\sum_{k=1}^N \langle 
|\vec{R}_k (t+tw)-\vec{R}_k(t_w)|^2\rangle;
\label{msd-eq}
\end{equation}
here $\langle \rangle$ is the average over the initial conditions. Note that 
in the calculations we have taken care of subtracting the displacement of 
the center of mass of the whole system, which does not vanish in Brownian 
simulations, to reveal the intrinsic dynamics of the particles. Indeed, 
in the case where the dynamics is very slow (as it is the case here), at 
long times the diffusion motion of the total center of mass can be 
relevant. In Fig.~\ref{fig:msd} we plot the mean-squared displacement at 
the indicated values of $t_w$. Following an initial transient regime, the 
curves reach a plateau, reminiscent of the cage effect in glassy systems,
and whose lifetime increases on increasing $t_w$; only at later times, 
eventually, the curves move from the plateau and apparently bend 
toward a constant value. From these data we can tentatively argue that, 
for values of $t_w$ longer than the one we can presently simulate, the 
system stops to flow on very long time scales and arrests in a disordered 
state. 

\begin{figure}[h]
\centering
\vspace{0.5cm}

\includegraphics[width=0.50\textwidth]{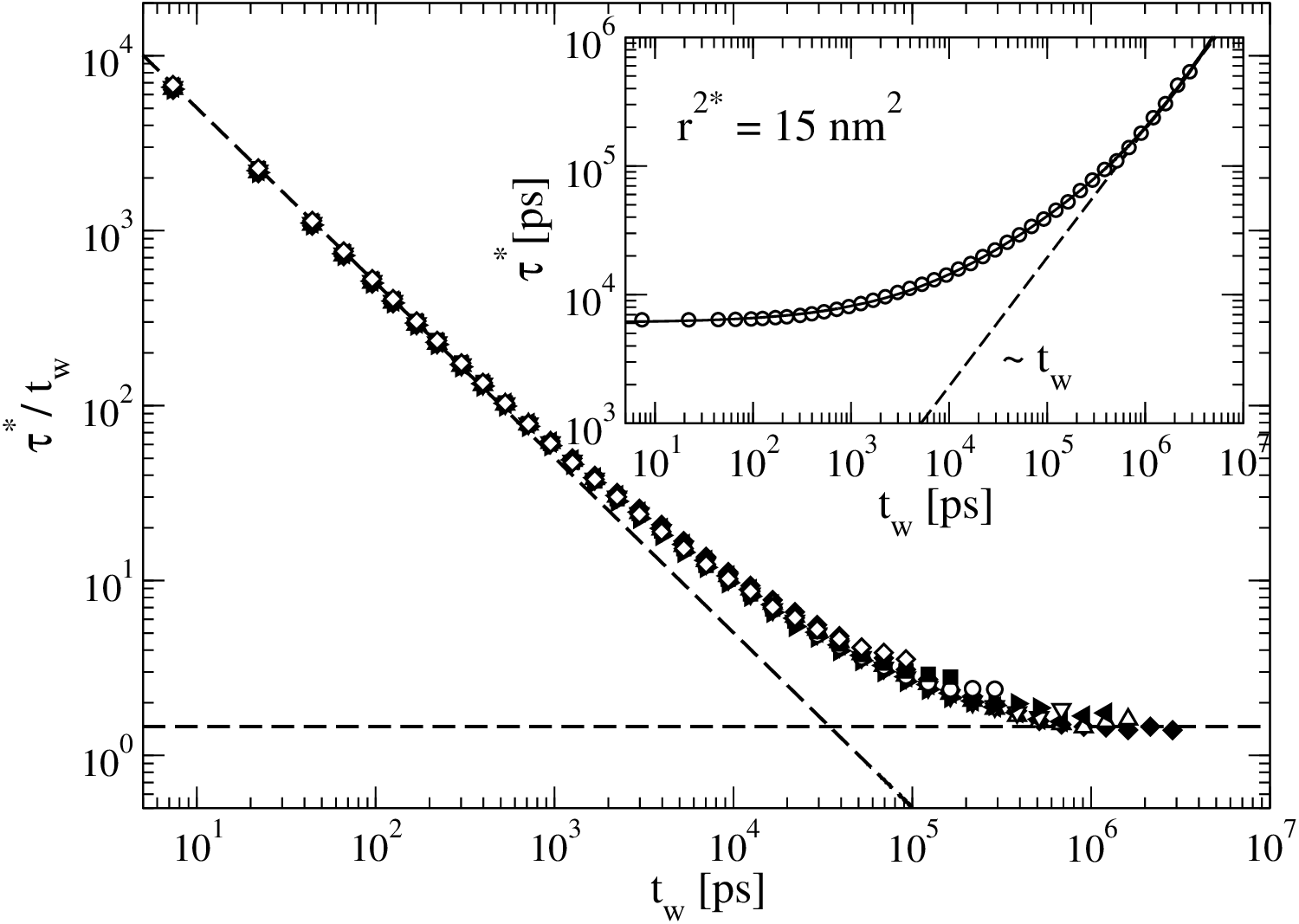}
\caption{}
\label{fig:taud}
\end{figure}

The absence of a clear diffusive regime (even at long times) and the absence 
of equilibrium conditions does not allow us to extract from the mean-squared 
displacement data a diffusion coefficient via the Einstein relation. In order 
to identify a relevant time scale of the structural dynamics, without 
introducing any ad-hoc model, we proceed  as follows: we fix an arbitrary 
threshold value $r^{2*}$, and consider the time $\tau^*$ at which 
$r^2(\tau^*)=r^{2*}$. We repeat this analysis as a function of the waiting 
time $t_w$. In Fig.~\ref{fig:taud} (main panel) we show our results for 
several values of $r^{2*}$, ranging from $5$ to $50$~nm$^2$. The data have 
been rescaled by $t_w$ and additionally shifted to maximize the overlap and 
stress the simple scaling with $t_w$ which, therefore, seems to set the only 
relevant time scale for diffusion. In the inset of Fig.~\ref{fig:taud} we 
show the raw data for the case $r^{2*}=15$~nm$^2$. At short $t_w$ the data 
can be represented by a function of the form $\tau^*\sim A + B t_w^x$,
with $x\simeq 0.64$. 
At longer $t_w$ we observe the linear behavior $\tau^*\propto t_w$, expected 
for $\tau^*\simeq t_w$. Indeed, from simple models (see, for instance, 
Ref.~\cite{bou92,bou00}), it is possible to show that the average life-time 
of a potential energy landscape basin, trapping the system at a time of order 
$t_w$, cannot exceed $t_w$. 

Already from these results it is clear that the aging dynamics forces us to
introduce a new relevant time scale, $t_w$. This is of particular interest
for the case of $Q$-dependent quantities, where we expect a particularly 
complex interplay between the length scales probed, the corresponding 
relaxation time $\tau_Q$ and the waiting time $t_w$. In the following 
Sections we therefore focus on the out-of-equilibrium dynamics of the 
density fluctuations.
\section{One-particle dynamics of density fluctuations}
\label{fqt-self:sec}
The relaxation dynamics of density fluctuations is encoded in the
intermediate scattering function $F(Q;t,t_w)=\langle\rho(Q,t)
\rho^*(Q,t_w)\rangle$, where the density fluctuations, $\rho(Q,t)$, 
have been defined above. (We recall here that $\langle\,\rangle$ is 
an average over the initial conditions which amounts to $100$ 
independent samples). In light scattering experiments in 
out-of-equilibrium conditions, information on the dynamics of the 
system is obtained by measuring the time autocorrelation function 
of the scattered intensity, 
$g_2(Q;t,t_w)=\langle I(t_w+t)I(t_w)\rangle$. 
In the single scattering approximation the time autocorrelation 
function of the scattered intensity and the intermediate scattering 
function are connected by the relation 
$F(Q;t,t_w)\propto\sqrt{g_2(Q;t,t_w)-1}$, where the proportionality 
factor depends on the detection set-up.

We start our study of the dynamics of density fluctuations focusing 
on the incoherent (one-particle) part of the intermediate scattering 
function, $F_s(Q;t,t_w)$,
\begin{equation}
F_s(Q;t,t_w)=\frac{1}{M}
\langle\sum_{i=1}^M e^{-i\vec Q\cdot\left[
{\vec X}_i(t+t_w)-{\vec X}_i(t_w)\right]}\rangle. 
\label{Fs:eq}
\end{equation}
We recall here that $\vec X_i$ can either refer to the center of 
mass of platelets $i$, in which case $M=N$, or sites positions, 
in which case $M=61\times N$. As already detailed above, the 
smallest value of the wave vector accessible in the simulation is 
$Q_m=2\pi/L\simeq 0.055$~nm$^{-1}$. We have used a wave vector 
resolution $\Delta Q=0.005$~nm$^{-1}$ and, for each value of $Q$, a 
maximum of $200$ $Q$-vectors has been considered in the calculation 
of averages.
We note that, although scattering experiments mainly probe the collective 
intermediate scattering function, we expect to grasp the main features 
of the long-time aging dynamics also from $F_s(Q;t,t_w)$. This choice is 
mainly due to the fact that one-particle dynamics is less affected by noise 
than collective dynamics and, therefore, statistics is more reliable in the 
former case. 

\begin{figure}[h]
\centering
\vspace{0.5cm}

\includegraphics[width=0.50\textwidth]{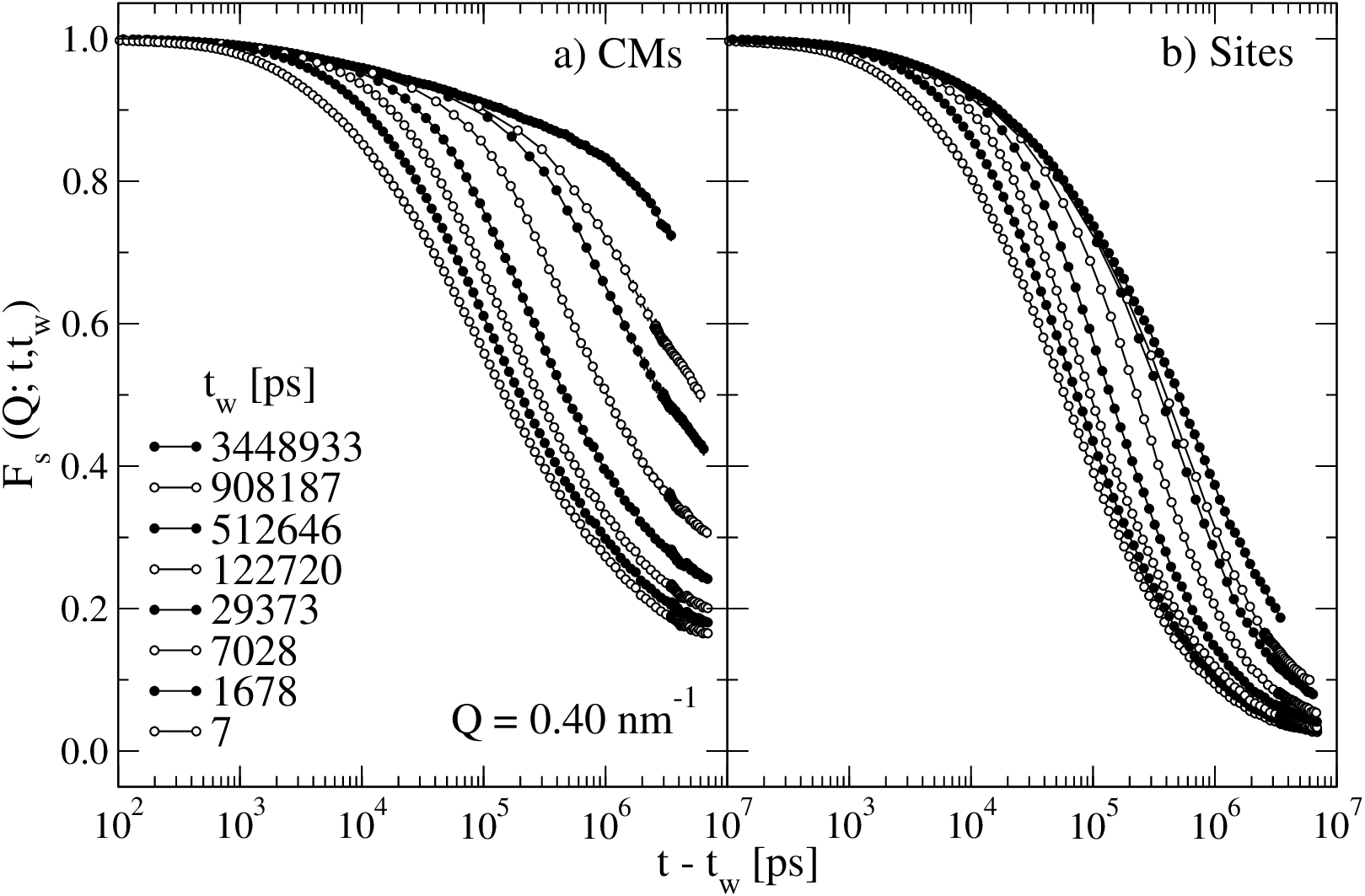}
\caption{}
\label{fig:self-tw}
\end{figure}

\begin{figure}[h]
\centering
\vspace{0.5cm}

\includegraphics[width=0.50\textwidth]{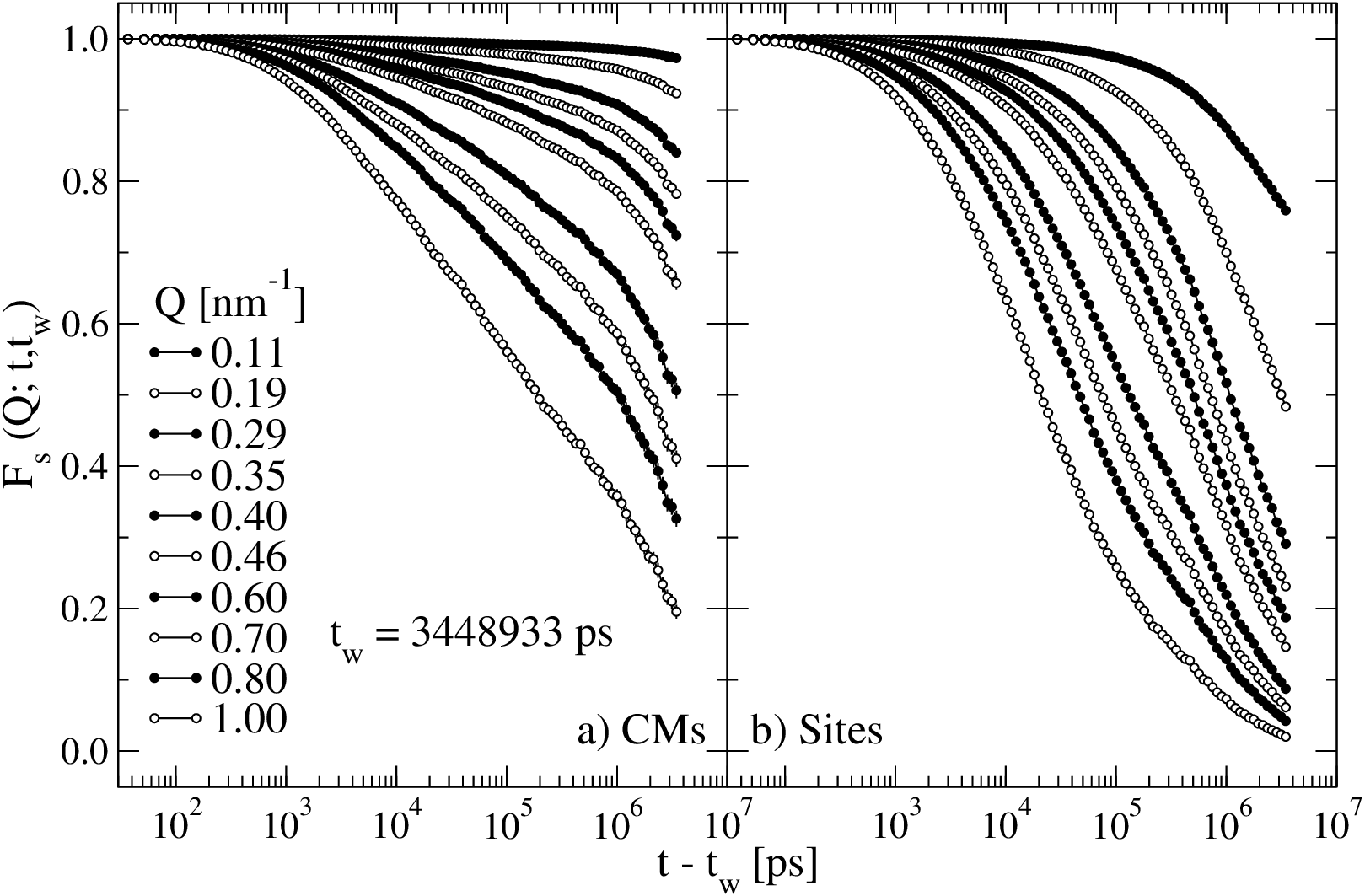}
\caption{}
\label{fig:self-Q}
\end{figure}

The quantities calculated from the position of the interaction 
sites also contain a dynamic contribution coming from rotations of the 
platelets; therefore, we expect the relative correlation functions to 
relax on time scales shorter than the analogous quantities calculated 
from the centers of mass of the platelets. Indeed, this is what we observe. 
In Fig.~\ref{fig:self-tw} we show our results at $Q=0.40$~nm$^{-1}$, at the 
indicated values of the waiting time $t_w$, both for centers of mass 
(left panel) and interaction sites (right panel). In Fig.~\ref{fig:self-Q} 
we show the simulation results at fixed $t_w\simeq 3.5 \mu$s at the 
indicated values of $Q$. From these two figures it is evident that aging 
is fully active in the system. In particular, the slowing down of the 
dynamics is larger the longer the aging time, and a clear $Q$-dependence
of the relaxation is visible at constant $t_w$. In the following we give 
a more detailed description of the effect of the aging time on the above
dynamics. 

The most natural choice would be to fit the data to a particular model and, 
then, to study the dependence on $Q$ and $t_w$ of the fitting parameters. 
In doing so we should take into account two important observations. 
First, we should note that most of the curves shown in Fig.~\ref{fig:self-tw}
and~\ref{fig:self-Q} do not relax to zero on the available time window and 
this fact severely limits the possibility to fit the data without 
introducing arbitrary bounds. Second, in the present case $t$ is always 
larger than or at most of the same order of magnitude of the waiting time 
$t_w$. This is in contrast with the case of experiments, where it is always 
$t_w \gg t$, and poses the question if it is reasonable to try to fit with 
a particular functional form a correlation function for times longer than the 
corresponding waiting time.

The above discussion amounts to identify a reasonable relaxation time scale 
for the intermediate scattering function which is independent of a particular 
model. We choose to use the method of the threshold we have described in the 
previous Section in the case of the mean squared displacement: 
we fix a particular threshold value and we consider as a relevant time scale 
the one at which the correlation function decreases from unity to the threshold 
value. Obviously, exploiting such a method we must take into account some caveats. 
Note, for instance, that the shape of the long time relaxation (possibly 
representable by a stretched exponential) clearly changes with the waiting 
time and, therefore, the relaxation time alone is not the only relevant parameter. 
Nevertheless, we expect to understand the correct general behavior and also 
clarify the interplay between the aging time and the length scales actually 
relevant for the aging dynamics. A-posteriori we also verify the above 
procedure in a particular case, fitting the data with a particular functional 
form and comparing the results.

\begin{figure}[h]
\centering
\vspace{0.5cm}

\includegraphics[width=0.50\textwidth]{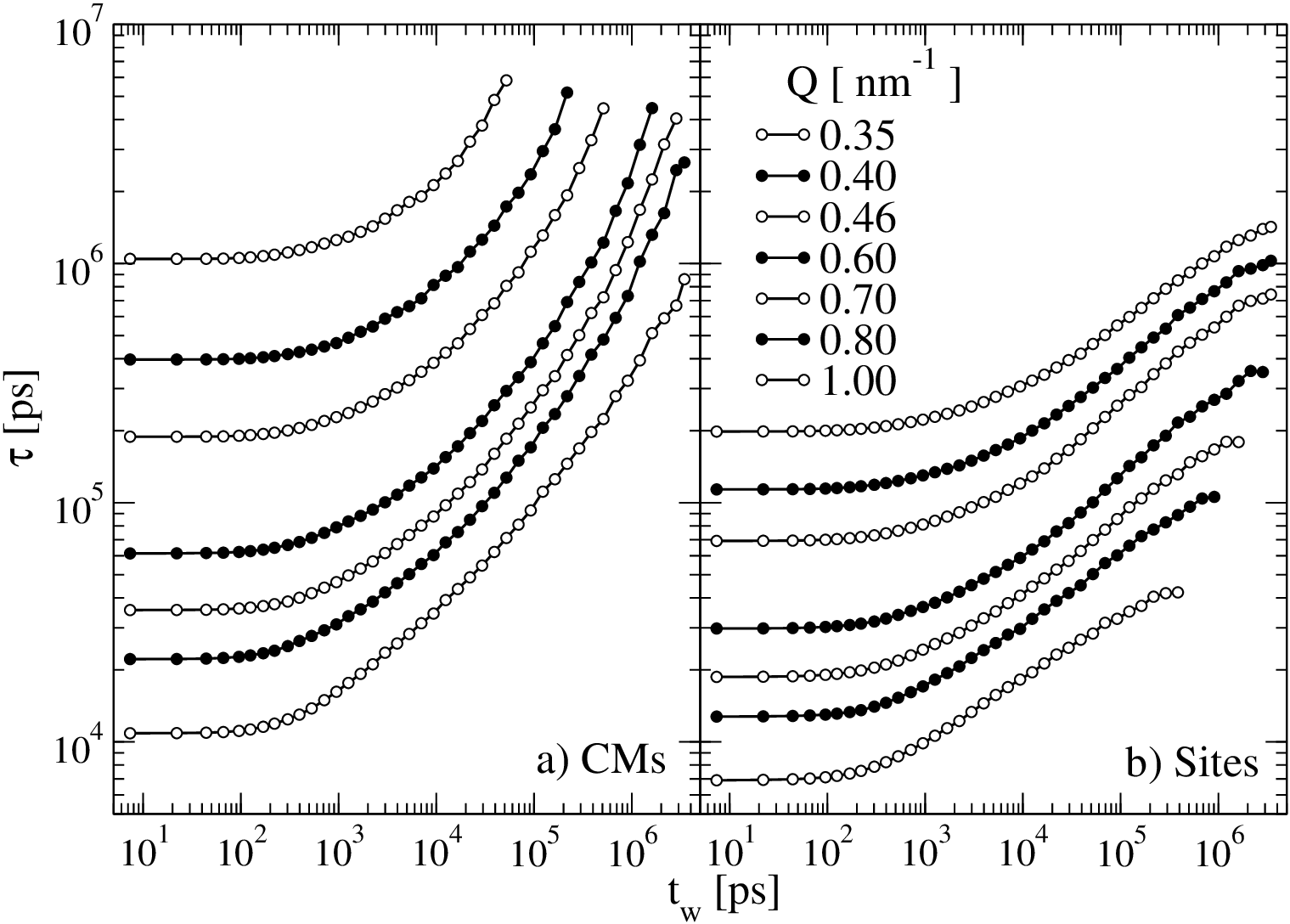}
\caption{}
\label{fig:tau-self-e}
\end{figure}

In Fig.~\ref{fig:tau-self-e} we show the results for both sites (left panel) 
and centers of mass (right panel) for a threshold value $f=e^{-1}$ as a 
function of the waiting time $t_w$, at the indicated values of $Q$. 
We have checked that the value chosen for the threshold does not 
qualitatively change the results. These data confirm the picture coming 
from the above analysis of the mean-squared displacement, pointing to a 
strong dependence of the out-of-equilibrium relaxation time on the waiting 
time. We observe that in our data is absent the exponential dependence 
on the waiting time of the relaxation time visible at short waiting times 
in experiments. (Note that the physical processes behind this particular
feature in experiments still remains unexplained~\cite{Tan05a}).

\begin{figure}[h]
\centering
\vspace{0.5cm}

\includegraphics[width=0.50\textwidth]{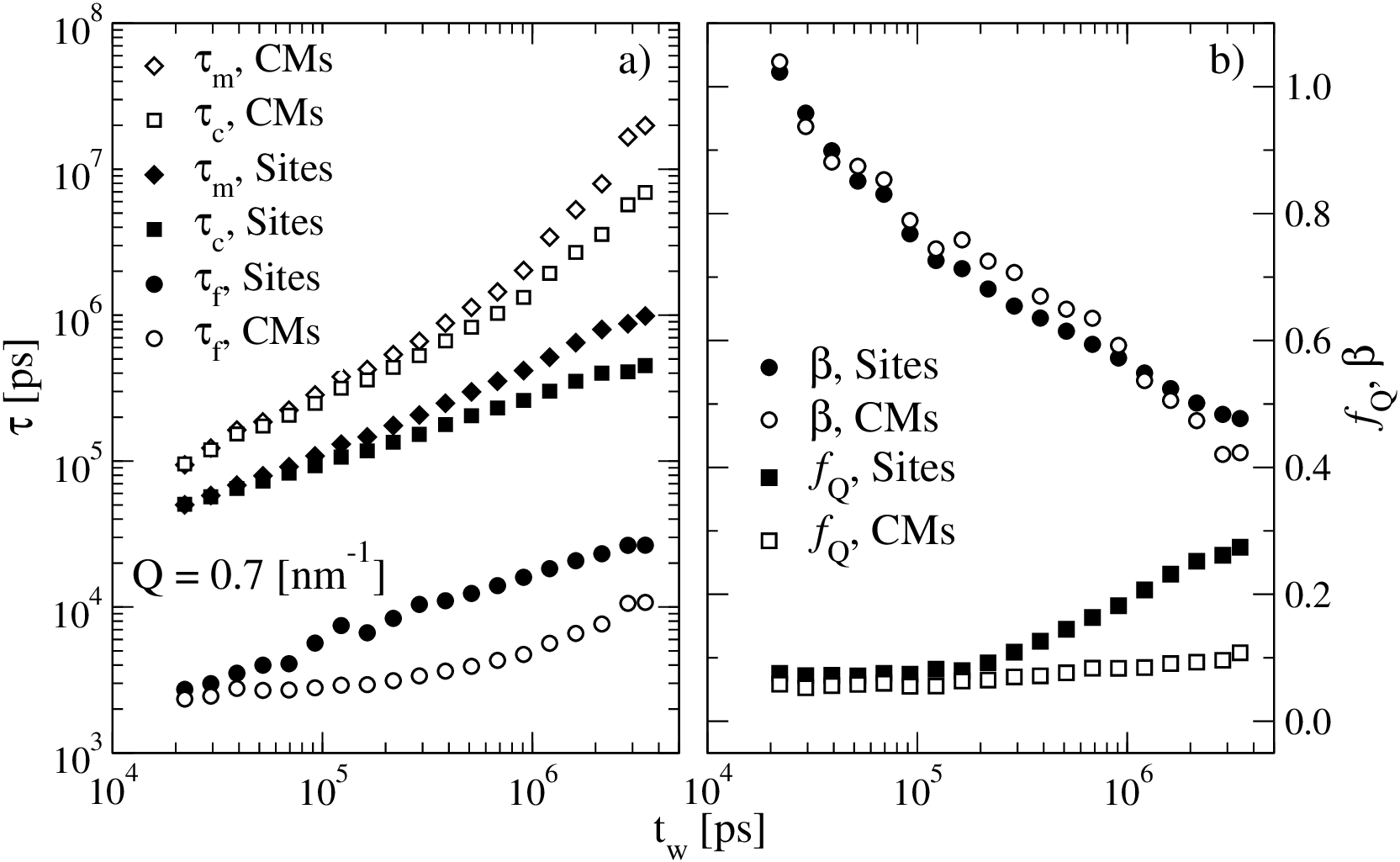}
\caption{}
\label{fig:fit-param-tw}
\end{figure}

\begin{figure}[h]
\centering
\vspace{0.5cm}

\includegraphics[width=0.50\textwidth]{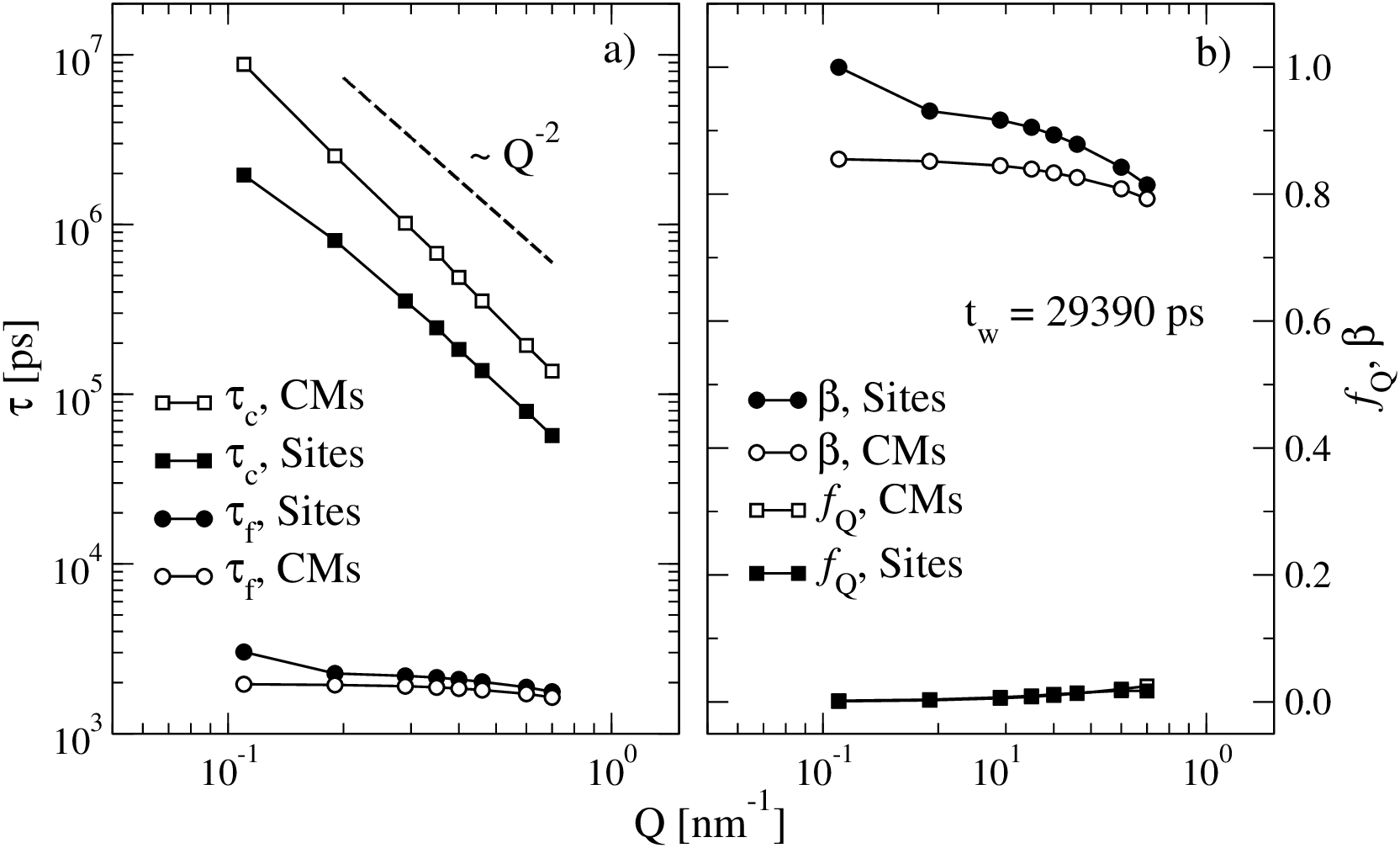}
\caption{}
\label{fig:fit-param-Q}
\end{figure}

To be more quantitative and check the analysis introduced above, we fit 
some of our data to a particular functional form, proposed some time ago 
and widely exploited in the literature (see, among others, 
Refs.~\cite{ruzicka04a,ruzicka04b,ruzicka06,abou01}). 
We choose, mainly due to a better quality of the data, the particular 
value $Q=0.7$ nm$^{-1}$ and fit the curves to the following model:
\begin{equation}
F_s(Q;t,t_w)=f_Q(t_w) e^{-t/\tau_f(Q;t_w)}
+(1-f_Q(t_w))e^{-(t/\tau_c(Q;t_w))^{\beta(Q;t_w)}}.
\label{fqt-fit}
\end{equation}
Here, we have introduced two relaxation times, i.e., $\tau_f$, related to 
fast short-times dynamics, and $\tau_c$, controlling the long-time 
relaxation. The non ergodicity parameter, $f_Q(t_w)$, and the stretching 
parameter, $\beta(Q;t_w)$, are reminiscent of the analogous quantities 
introduced for the study of the glass transition. We fit our data to 
Eq.~(\ref{fqt-fit}) only for times tentatively shorter than $2\: t_w$. 
Fig.~\ref{fig:fit-param-tw} shows the $t_w$ dependence of fitting 
parameters. The left panel shows $\tau_c$, $\tau_f$ and the ``mean'' 
relaxation time $\tau_m$, defined as $\tau_m = \tau_c \Gamma(1/\beta)/\beta$, 
both for centers of mass of the platelets and interaction sites.
All the data strongly depends on the waiting time and increase with $t_w$. 
The right panel shows our results for $\beta$, decreasing with $t_w$ 
from $1$ to about $0.4$  (this behavior is also observed in experiments, 
see~\cite{abou01}), and $f_Q$, which stays constant at a value of about 
$0.1$ in the case of the center of mass and increases up to $0.3$ for sites. 
Fig.~\ref{fig:fit-param-Q} shows the $Q$-dependence of the same fitting 
parameters at constant $t_w \simeq 30$ ns. $\tau_f$ stays almost constant 
in the whole investigated $Q$-range, while $\tau_c$ shows the typical 
$Q^{-2}$ dependence~\cite{schosseler06,bellour03,cipelletti03}. 
Also, $f_Q$ stays almost constant in whole considered range (data for
platelets and sites overlap almost perfectly), while $\beta$ shows a 
slight decrease with $Q$ in both cases.
 
We note that the above fits are particularly difficult due to the low 
statistics and to the fact that not all the curves relax to zero on the 
accessible time window. Moreover, the absence of a clear separation 
between time scales, makes particularly difficult the determination of 
the relaxation times. More specifically, in some cases it was also 
possible to fit the data with two comparable values for $\tau_f$ and 
$\tau_c$, obtaining a value of $\chi^2$ comparable to the one of the 
actual fit with $\tau_f \ll \tau_c$.
\section{Collective dynamics of density fluctuations}
\label{fqt-coll:sec}
\begin{figure}[h]
\centering
\vspace{0.5cm}

\includegraphics[width=0.50\textwidth]{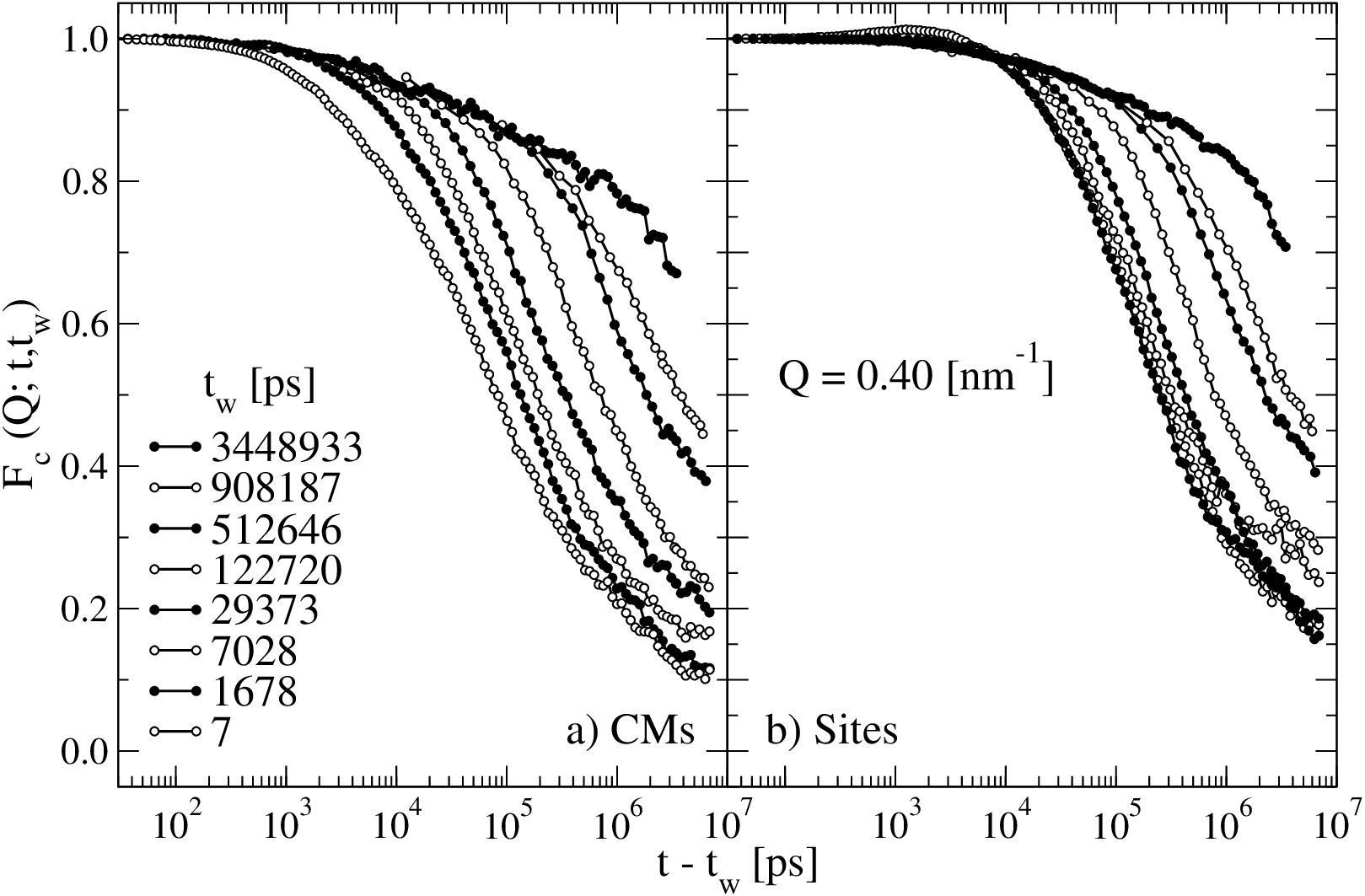}
\caption{}
\label{fig:coll-tw}
\end{figure}

\begin{figure}[h]
\centering
\vspace{0.5cm}

\includegraphics[width=0.50\textwidth]{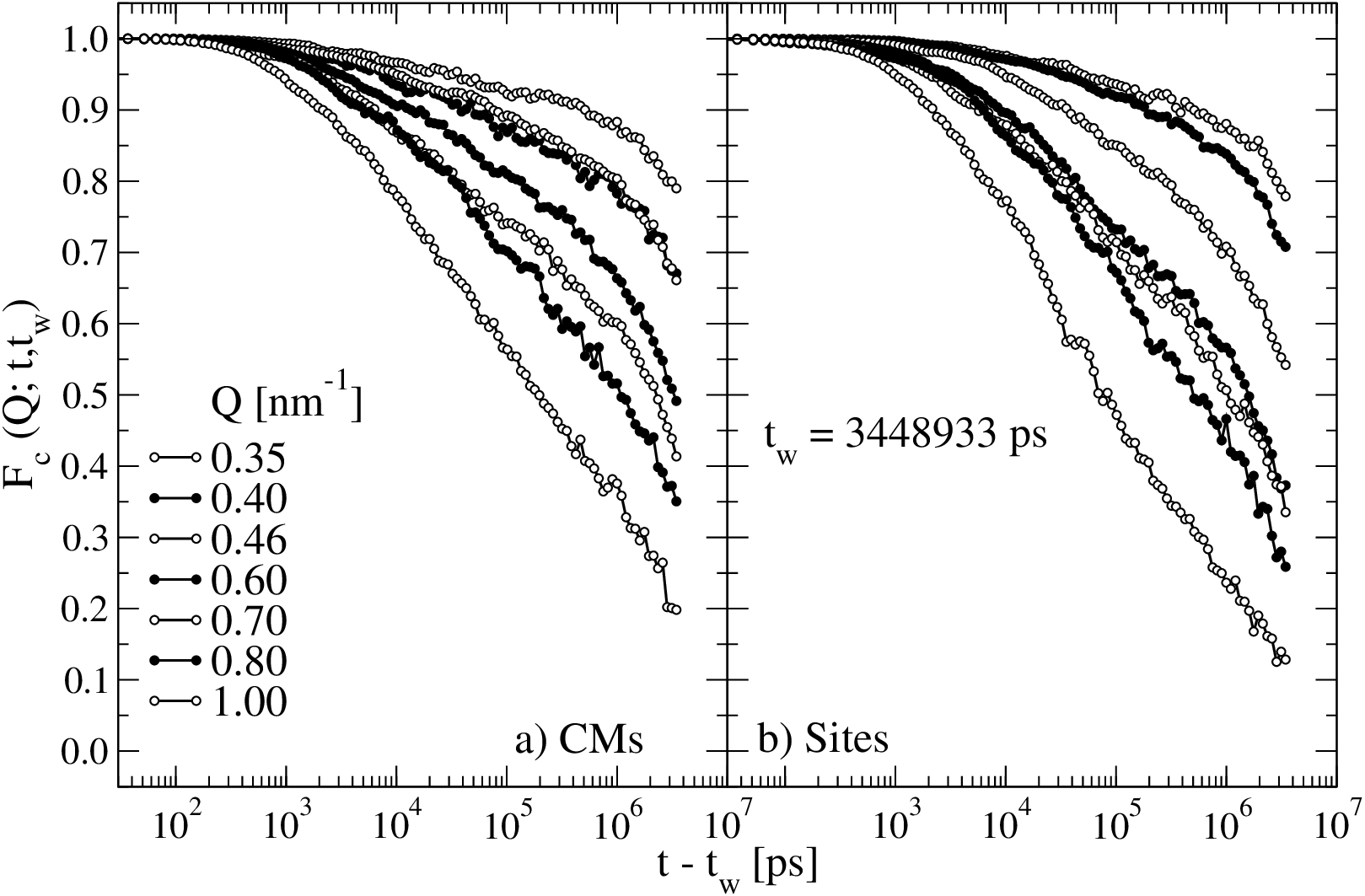}
\caption{}
\label{fig:coll-Q}
\end{figure}

The coherent intermediate scattering function at times $t_w$ and $t$ is 
defined as
\begin{equation}
F_c(Q;t,t_w)=\frac{1}{M S(Q)}\langle\sum_{i=1}^M \sum_{j=1}^M 
e^{-i\mathbf{Q}\cdot\left[\mathbf{X}_i(t+t_w)-\mathbf{X}_j(t_w)\right]}\rangle,
\label{Fc:eq}
\end{equation}
where $X_i$ has been defined above and $S(Q)$ are the data included in 
Fig.~\ref{fig:sk}. This function is related to the collective dynamics 
of the density fluctuations, and it is the quantity of direct relevance
for the experimental measurements. 
In Fig.~\ref{fig:coll-tw} we show the $t_w$-dependence of 
$F_c(Q,t)$ at $Q=0.40$ nm$^{-1}$, both for centers of mass (left panel)
and sites (right panel), while in Fig.~\ref{fig:coll-Q} we show the 
$Q$-dependence of $F_c(Q,t)$ at fixed $t_w\simeq 3.4 \:\mu$s,
again for the two cases. Also these results confirm the strong 
$t_w$-dependence of the out-of-equilibrium dynamics and do not seem
to add much more insight to the analysis of the previous Section. 
Indeed, the qualitative behavior of $F_c(Q; t, t_w)$ and $F_s(Q; t, t_w)$, 
as a function of both $Q$ or $t_w$, is very similar, as it is evident by 
comparing Fig.~\ref{fig:coll-tw} to Fig.~\ref{fig:self-tw} and 
Fig.~\ref{fig:coll-Q} to Fig.~\ref{fig:self-Q}. Hence, the same conclusions 
of Sec.~\ref{fqt-self:sec} can also be drawn from these results, at least
at a qualitative level. 

\begin{figure}[h]
\centering
\vspace{0.5cm}

\includegraphics[width=0.50\textwidth]{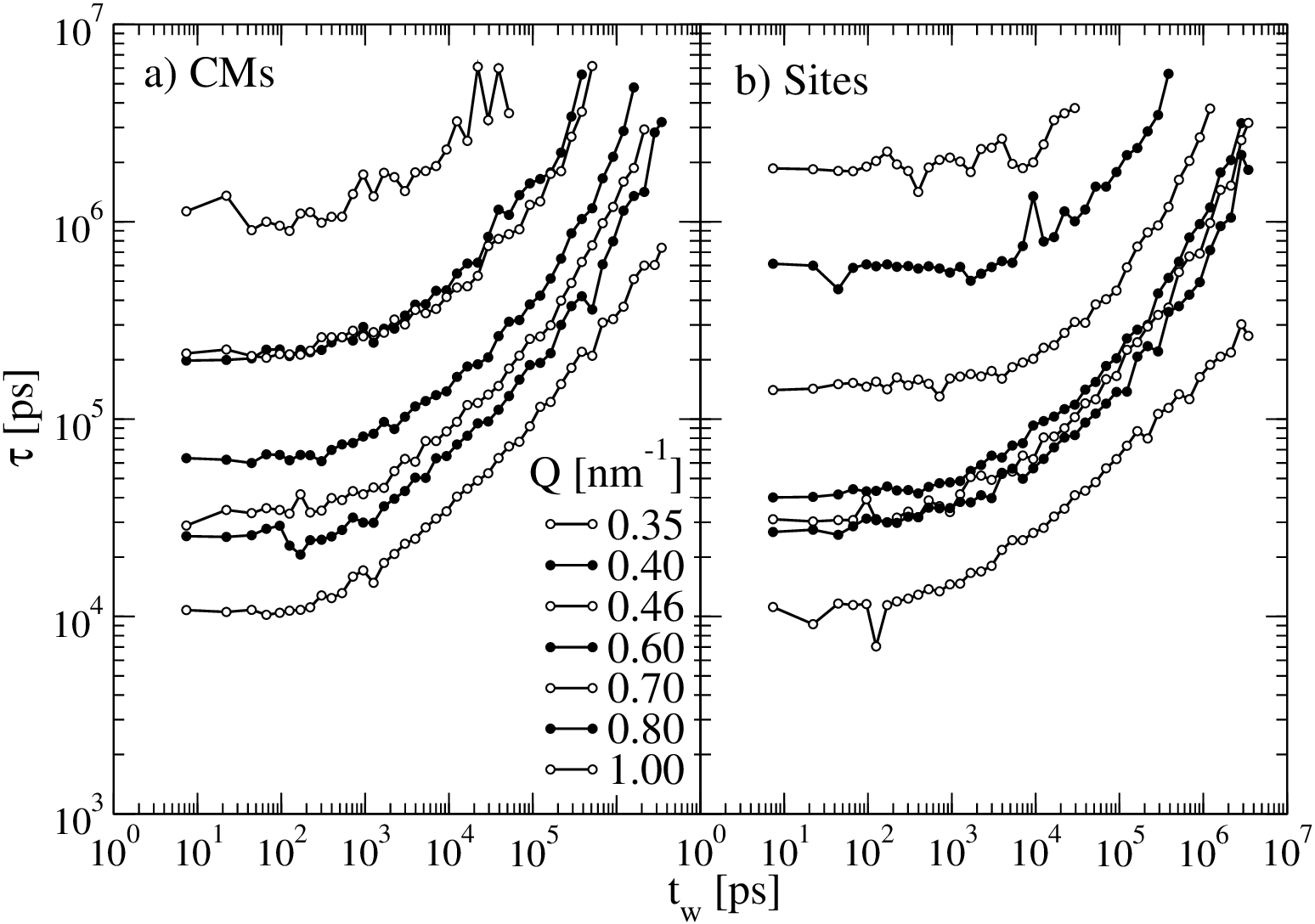}
\caption{}
\label{fig:tau-coll-e}
\end{figure}

In the present case we do not attempt a direct fit of the data and directly
exploit the method of the threshold to evaluate a reasonable relaxation time 
scale, independent of a particular functional form for a fitting model. 
Fig.~\ref{fig:tau-coll-e} shows our results, with a threshold $f=e^{-1}$, 
for the data at the indicated values of $Q$ and as a function of the waiting 
time, $t_w$. Again, we have checked that the results do not qualitatively 
change with the chosen value of the threshold. Data are qualitatively in 
agreement with the results reported in Fig.~\ref{fig:tau-self-e} and 
confirm the overall picture we described in details in the previous Section.
\section{Rotational Dynamics}
\label{rotations:sec}
The strong anisotropy of the Laponite molecule implies that an important
role in dynamics and structural reorganization must be played by the mutual 
orientations of the platelets. Therefore, also from the above results, we 
expect that the orientational behavior of the system, both for the single 
platelet and collective, must strongly couple to the dynamics of density 
fluctuations and take place on comparable time scales. 
This is the main motivation for the present Section.

The behaviour of the orientational degrees of freedom is accessible via 
the dynamics of the Legendre polynomials calculated from the orientation 
vectors of platelets. Some of those polynomial are directly related to 
the response functions detected by different experimental techniques. 
The convenient correlation functions in this case are the functions
$C_l(t,t_w)$, and their self-part, $C_l^{(s)}(t,t_w)$, which are 
defined as:
\begin{subequations}
\begin{eqnarray}
C_l(t,t_w)&=&\frac{1}{N}\langle\sum_{i=1}^N \sum_{j=1}^N
P_l(\vec{u}_i(t+t_w)\cdot\vec{u}_j(t_w))\rangle\\
C_l^{(s)}(t,t_w)&=&\frac{1}{N}\langle\sum_{i=1}^N
P_l(\vec{u}_i(t+t_w)\cdot\vec{u}_i(t_w))\rangle.
\label{legendre1}
\end{eqnarray}
\end{subequations}
Here, ${\vec u}_i$ is the normalized orientational vector of the $i$~th 
molecule, pointing along the normal to the surface of the platelet, and 
$P_l(x)$ is the Legendre polynomial of order $l$, with $l\ge 1$. 
Note that for $l=1$ and $l=2$ the functions $C_l(t)$ can be measured in 
dielectric and light scattering experiments. We also note that it is 
often assumed that the cross-term in $C_l(t)$ can be neglected; in this 
case the experiments would also yield information on $C_l^{(s)}(t)$.

\begin{figure}[h]
\centering
\vspace{0.5cm}

\includegraphics[width=0.50\textwidth]{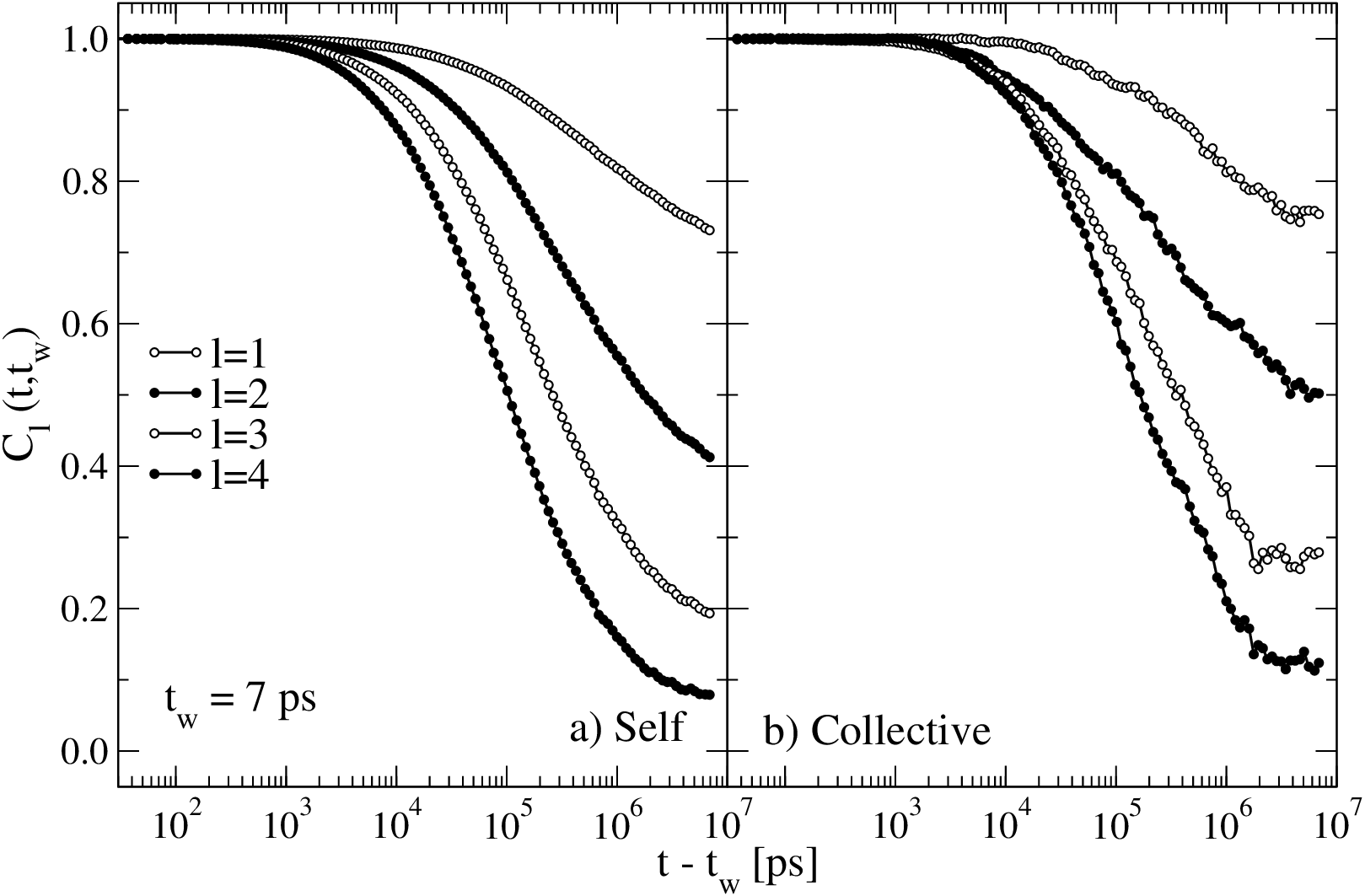}
\caption{}
\label{fig:rot-l}
\end{figure}

In Fig.~\ref{fig:rot-l} we show our results at $t_w=7$~ps at the 
indicated values of $1\le l\le 4$, for the one-particle (left panel) 
and collective (right panel) cases respectively. From these data it 
is evident that the smaller the value of $l$, the longer the time 
scale of the relaxation dynamics. In particular, note that in the 
collective case, the curves seem to saturate to a finite value at 
long time, pointing to a progressive freezing of the orientational
degrees of freedom at very long times. Unfortunately the limited 
total time extension of the present computer simulation  does not 
allow us to be conclusive on this point. 

\begin{figure}[h]
\centering
\vspace{0.5cm}

\includegraphics[width=0.50\textwidth]{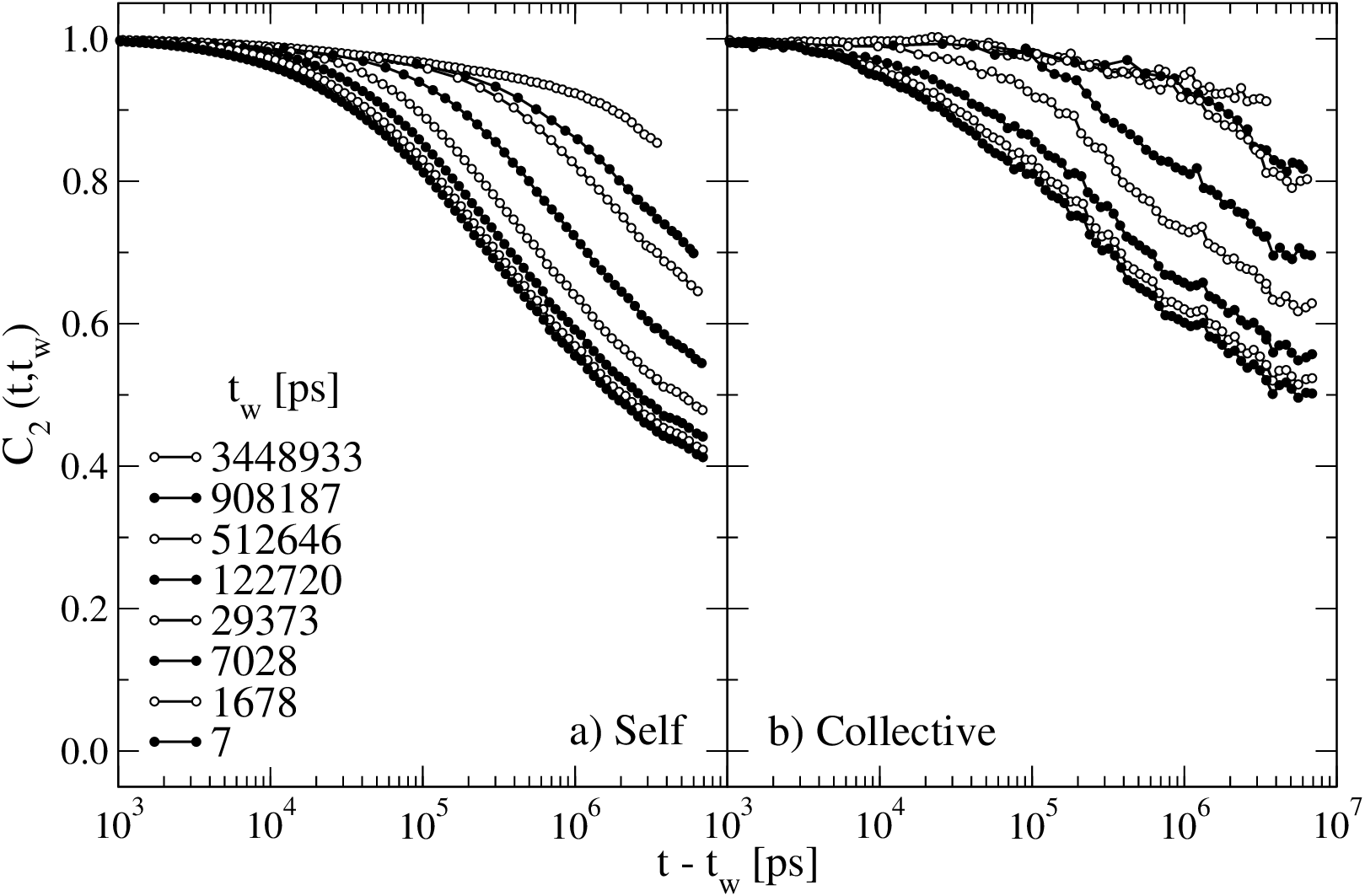}

\vspace{1.0cm}

\includegraphics[width=0.50\textwidth]{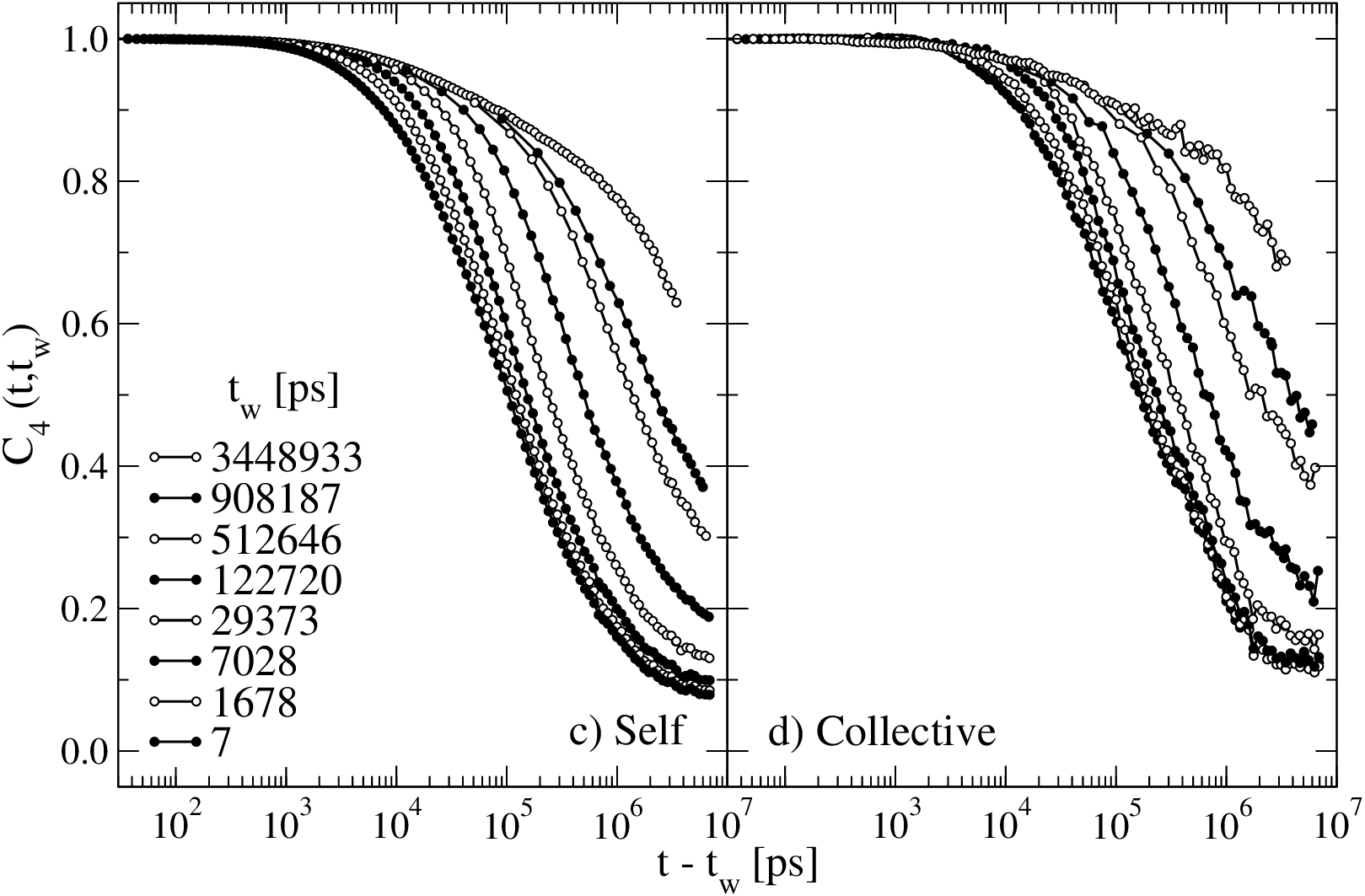}
\caption{}
\label{fig:rot-tw}
\end{figure}

In Fig.~\ref{fig:rot-tw} we plot our results for $l=2$ 
(panels a) and b)) and $4$ (panel c) and d)) at the indicated 
waiting times, both for the self (left panels) and collective 
(right panels) cases. Again, also in the case of the rotational 
dynamics, a strong aging time dependence is evident and at longer 
waiting times the rotational dynamics slows down significantly. 

\begin{figure}[h]
\centering
\vspace{0.5cm}

\includegraphics[width=0.50\textwidth]{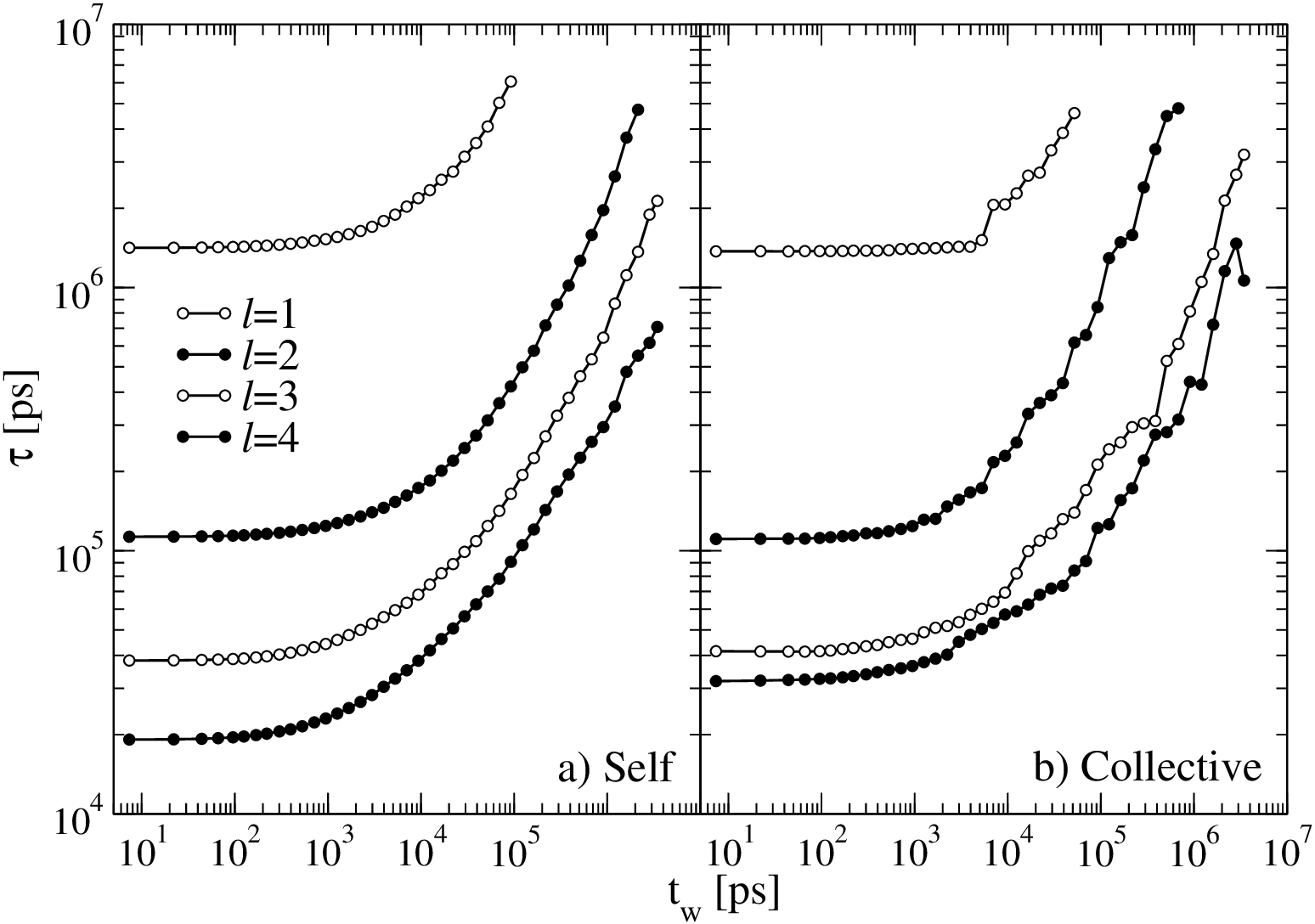}
\caption{}
\label{fig:rot-tau}
\end{figure}

A threshold analysis similar to the one described above for the dynamics 
of density fluctuations allows one to identify a relaxation time. Our 
results are shown for $1\le l \le 4$ in Fig.~\ref{fig:rot-tau}, where 
the threshold has been fixed to a value $f=0.8$. The overall qualitative 
behavior is analogous to the one found in the case of the translational 
dynamics, and the time scales, except for the case $l=1$, are also 
comparable in the two cases. Concluding, our data support a strong 
coupling between translational and orientational degrees of 
freedom~\cite{jabbari-farouji04,fabbian98}, as expected for a strongly
anisotropic molecule.
\section{Conclusions}
\label{conclusions:sec}
In this work we have presented a systematic study of the out-of-equilibrium 
and aging dynamics of a model for a Laponite colloidal suspension. The 
choice of the system considered has been mainly suggested by the terrific 
amount of experimental work already present in the literature. 
A realistic site-site, purely repulsive Yukawa-like interaction potential 
between Laponite platelets, has been considered. We have used the Brownian 
dynamics technique for extensive molecular dynamics simulations. 
In particular, we have proposed a new algorithm which, although neglecting 
hydrodynamic effects, correctly takes into account the strong anisotropy 
of the Laponite platelet, introducing two different friction coefficients, 
respectively parallel and perpendicular to the direction of the symmetry 
axis of the platelet.

An extensive ensemble of independent systems has been driven 
out-of-equilibrium by an instantaneous jump from a high temperature to room 
temperature at constant volume fraction and realistic configurations have 
been stored for the analysis. In particular, we have focused on the 
long-time aging dynamics, calculating the mean-squared displacement and
the intermediate scattering functions. Both coherent and incoherent dynamics 
have been studied in details. The dynamics of the density fluctuations has 
been found to be strongly dependent on the aging time and a qualitative 
discussion has been performed on the interplay between the different time 
and length scales which play a significant role in the relaxation. 
A study of the rotational dynamics has completed the study of the dynamics 
of the Laponite colloidal suspension. The main result is that aging dynamics 
also strongly affects the orientational degrees of freedom, which relax on 
time scales comparable to the ones typical of translational modes.  

The study presented here strongly suggests that, indeed, in the absence of
attractive interactions between molecules, a disordered arrested state can 
be generated by the only presence of Yukawa repulsive interactions. 
The long range of the repulsive interaction makes it possible to arrest the 
dynamics of the colloidal suspension even if the effective packing fraction 
is only a few percent. In this respect, cages are defined not by the physical 
size of the hard-core but by the effective range of the screened electrostatic 
potential. While in the case of spherical particles, Yukawa interactions 
alone hardly generate a glass, due to the fast rates of crystallization 
induced by the long range of the interaction, the anisotropy of the platelets 
appears to be able to self-generate a sufficient local disorder which 
stabilizes the metastability of the arrested disordered state. 

In this respect, despite our data are not conclusive, due to the difficulty 
of extending the time length of the simulation, charged platelets may give 
rise to Wigner glasses. Future work, in line with the present approach, 
must focus on the role played in the dynamics by both the salt concentration 
and the presence of opposite sign rim charges. 
\newpage
\bibliographystyle{./apsrev}
\bibliography{./paper_bib}
\newpage
\noindent{\bf FIGURE CAPTIONS}
\vspace{1.5cm}

\noindent{\bf FIG 1:} Relaxation of potential energy as a function of time 
after the temperature jump. $100$ independent realizations of the system 
have been equilibrated at high temperature, $T=500$~K, and, at $t_w=0$,
instantaneously quenched to $T=300$~K. In the plot we show the data for 
each run (lines) and the averaged value (symbols). Note that the energy 
stays positive, due to the purely repulsive nature of the interaction 
site-site potential. The relaxation of the energy follows a power law 
behavior of exponent $\gamma\simeq 0.35$ and the value predicted at 
equilibrium (estimated at $E_{\infty} \simeq 206.6$~eV) is not reached 
on the total simulation time scale. Therefore, the systems stay 
out-of-equilibrium for the total duration of the present simulation.

\noindent{\bf FIG 2:} {\em Main Panel}: Equilibrium static structure 
factor, $S(Q)$, calculated both for the platelets centers of mass 
(open circles) and interaction sites (closed circles). Each point 
is an average over $100$ system configurations for $t>10^6$~ps. 
The smallest value of $Q$ accessible in our simulation is 
$Q_m=2\pi/L\simeq 0.05$~nm$^{-1}$. For the wave-vector average, we 
have chosen a resolution $\Delta Q=0.005$~nm$^{-1}$ and up to 
$200$ $Q$-vectors have been considered for each value of $Q$. 
Error bars are smaller than the symbols. {\em Inset}: Variation of 
the static structure factor calculated on sites at short times at 
the indicated values of $t$. A slight change in the structure of the 
second sub-peak is evident.

\noindent{\bf FIG 3:} Mean squared displacement $\langle r^2(t, t_w)\rangle$, 
calculated for the platelets centers of mass, as discussed in the text, as 
a function of time for different waiting times $t_w$. Shorter waiting 
times are on the top. The initial transient dynamics at short $t$ is 
followed by a plateau whose life-time increases on increasing $t_w$. 
Next the curves move from the plateau, saturating to a (nearly) constant 
value at very long times. From these data we could argue that the system 
eventually stops to flow on very long time scales.

\noindent{\bf FIG 4:} Relevant dynamic time scale $\tau^*$ calculated from 
the mean squared displacement, as discussed in the text. {\em Main panel:} 
$\tau^*$ as a function of $t_w$ calculated for different values of $r^{2*}$, 
ranging from $5$ to $50$~nm$^2$. Data have been rescaled by $t_w$ and 
additionally shifted to maximize the overlap and stress the simple scaling 
with $t_w$, which sets the relevant time scale. {\em Inset:} $\tau^*$ as a 
function of $t_w$ for $r^{2*}=15$~nm$^2$. At short $t_w$ data can be 
represented by a power law behavior, $\tau^* \simeq A + B t_w^x$, with 
$x \simeq 0.64$ (solid line); at longer $t_w$, one observes the expected 
linear behavior $\tau^*\propto t_w$ (dashed line)~\cite{kaloun05b}.

\noindent{\bf FIG 5:} $t_w$-dependence of the incoherent (one-particle) 
intermediate scattering functions $F_s(Q;t,t_w)$ at $Q=0.40$~nm$^{-1}$, 
at the indicated values of $t_w$. {\em a)} Data calculated from the
positions of the platelets centers of mass. {\em b)} Data calculated 
from sites, also containing orientational dynamics contributions. 
The effect of aging is evident from these data.

\noindent{\bf FIG 6:} $Q$-dependence of the one-particle intermediate 
scattering function $F_s(Q;t,t_w)$ at $t_w=3448933$~ps at the indicated 
values of $Q$. {\em a)} Data calculated on the platelets centers of mass. 
{\em b)} Data calculated on interaction sites. A strong dependence on $Q$ 
is evident from these data.

\noindent{\bf FIG 7:} Relaxation time for the incoherent intermediate 
scattering functions calculated by the threshold method with $f=e^{-1}$ 
for both centers of mass and sites as a function of the waiting time 
$t_w$. The indicated values of $Q$ have been considered.

\noindent{\bf FIG 8:} Waiting time dependence of the parameters of the fit 
discussed in the text for the incoherent scattering function at 
$Q=0.7$~nm$^{-1}$.

\noindent{\bf FIG 9:} Parameters of the fits discussed in the text for 
the incoherent scattering function at $t_w = 29390$~ps as a function 
of the wave vector $Q$.

\noindent{\bf FIG 10:} $t_w$-dependence of the coherent (collective) 
intermediate scattering function, $F_c(Q;t,t_w)$, at $Q=0.40$ nm$^{-1}$, 
at the indicated values of $t_w$ (longer $t_w$'s on top). Data calculated on 
the platelets centers of mass ({\em a)}) and on interaction sites ({\em b)}) 
are qualitatively in agreement.

\noindent{\bf FIG 11:} $Q-$ dependence of the coherent intermediate scattering 
function, $F_c(Q;t,t_w)$, at $t_w=3448933$~ps at the indicated values of $Q$. 
Data calculated on the platelets centers of mass ({\em a)}) and on interaction 
sites ({\em b)}) show a qualitative similar behavior.

\noindent{\bf FIG 12:} Relaxation time for the coherent intermediate 
scattering functions calculated by the threshold method with $f=e^{-1}$ 
for centers of mass and sites as a function of the waiting time $t_w$ 
at the indicated values of $Q$.

\noindent{\bf FIG 13:} Correlation functions of the Legendre polynomials 
of the order $1\le l \le 4$ at $t_w\simeq7$~ps. We show the self 
contribution (left panel) and collective total case (right panel). 
The orientational relaxation dynamics is slower the higher the value 
of the order $l$.

\noindent{\bf FIG 14:} {\em Top}: Waiting time dependence of the orientational 
correlation functions for $l=2$ (longer waiting times on the top), both for 
the self (left panel) and collective (right panel) cases.
{\em Bottom}: Waiting time dependence of the orientational correlation 
functions for $l=4$ (shorter waiting times on the top) both for the 
self (left panel) and collective total case (right panel).

\noindent{\bf FIG 15:} Aging time dependence of the orientational relaxation 
time for the correlation functions of the Legendre polynomials of order 
$1\le l \le 4$ determined by means of the threshold method, as discussed in 
the text. The threshold has been fixed to the value of $f=0.8$. 
{\em Left panel}: Data calculated from the self part of the correlation 
functions; {\em Right panel}: Data calculated from the total collective 
correlation functions.
\newpage
\vspace{10.0cm}

\newpage
\vspace{3.0cm}

\newpage
\newpage
\newpage
\newpage
\newpage
\newpage
\newpage
\newpage
\newpage
\newpage
\newpage
\newpage
\newpage
\newpage
\end{document}